\documentclass[%
 amsmath,amssymb,
preprint,%
]{revtex4-2}

\usepackage{graphicx}
\usepackage{dcolumn}
\usepackage{bm}
\usepackage{float}

\usepackage[utf8]{inputenc}
\usepackage[T1]{fontenc}
\usepackage{mathptmx}
\usepackage{etoolbox}
\usepackage[normalem]{ulem}

\usepackage{xcolor} 

\makeatletter
\def\@email#1#2{%
 \endgroup
 \patchcmd{\titleblock@produce}
  {\frontmatter@RRAPformat}
  {\frontmatter@RRAPformat{\produce@RRAP{*#1\href{mailto:#2}{#2}}}\frontmatter@RRAPformat}
  {}{}
}%
\makeatother
\begin{document}


\author{Sivanandan Kavuri$^1$, Chander Shekhar Sharma$^2$, George Karapetsas$^3$ and Kirti Chandra Sahu$^1$}
\affiliation{$^1$Department of Chemical Engineering, Indian Institute of Technology Hyderabad, Kandi - 502284, Telangana, India
\\
$^2$Thermofluidics Research Laboratory, Department of Mechanical Engineering, Indian Institute of Technology Ropar, 140001 Rupnagar, India\\
$^3$Department of Chemical Engineering, Aristotle University of Thessaloniki, Thessaloniki 54124, Greece}
\email{chander.sharma@iitrpr.ac.in, gkarapetsas@auth.gr, ksahu@che.iith.ac.in}
\title{Evaporation of sessile drops on a heated superhydrophobic substrate}

\date{\today}

\begin{abstract}
We experimentally investigate the evaporation dynamics of sessile water droplets on a micro-nano textured superhydrophobic aluminum substrate at various temperatures using shadowgraphy imaging. By comparing the evaporation behavior of two droplets placed side-by-side with that of an isolated droplet, we find that droplets in the two-drop system evaporate more slowly due to the vapor shielding effect, which increases vapor concentration between the droplets. This leads to asymmetric evaporation and longer lifetimes, particularly at room temperature compared to higher temperatures. At room temperature, the isolated droplet primarily follows a constant contact angle (CCA) mode, with occasional stick-slip events. The two-drop system predominantly exhibits CCA mode for most of its lifetime, with mixed-mode evaporation and some stick-slip behavior. At elevated temperatures, the isolated droplet maintains a nearly constant contact angle for the first half of its lifetime, transitioning to a mixed evaporation mode with occasional stick-slip events. In contrast, the two-drop system follows a mixed evaporation mode throughout, with occasional stick-slip behavior. Furthermore, a comprehensive theoretical model that accounts for diffusion, evaporative cooling, and convection accurately captures the evaporation dynamics of sessile droplets on a superhydrophobic substrate in both isolated and paired configurations at elevated substrate temperatures. In contrast, a diffusion-based model alone adequately describes the evaporation behaviour at room temperature.
\end{abstract}

\maketitle


\section{Introduction}
\label{sec:intro}

The evaporation of sessile drops is crucial for a variety of applications, including biomedical fields \cite{garcia2017sessile, gelderblom2022evaporation,balusamy2021lifetime,katre2021fluid}, electronics \cite{bar2006direct}, optics \cite{takamatsu2023all}, crystallization \cite{zang2019evaporation}, hot spot cooling \cite{kumari2011characterization,kim2007spray}, thin film devices \cite{prasad2014monitoring}, nanomaterial synthesis \cite{bridonneau2020self}, inkjet printing \cite{lohse2022fundamental,de2004inkjet}, and pharmaceuticals \cite{christodoulou2017modelling}, to name a few. While much of the previous research has focused on the evaporation of isolated sessile droplets, real-world scenarios often involve arrays or randomly distributed droplets on surfaces. In such multi-drop configurations, the presence of neighbouring droplets reduces evaporation rates by increasing local humidity and creating a shielding effect \cite{schofield2020shielding}. To understand these intriguing phenomena, a simplified approach involves comparing the evaporation dynamics of isolated droplets with those of a two-drop configuration that can offer valuable insights into the underlying physical processes. 

Previous studies have extensively investigated the evaporation dynamics of multiple-drop configurations on hydrophilic and hydrophobic substrates under both room temperature conditions \cite{iqtidar2023drying, masoud2021evaporation, wray2020competitive, Malachtari_interaction_2024} and elevated substrate temperatures \cite{Hari2024}. In contrast, the behavior of sessile droplets on superhydrophobic surfaces has received comparatively limited attention \cite{erbil2023droplet, moradi2020collective}, with most investigations concentrating on room temperature scenarios. This is despite the critical role superhydrophobic surfaces play in various applications, including the manufacturing of waterproof fabrics \cite{roach2008progess}, reducing underwater drag \cite{nosonovsky2009superhydrophobic, bhushan2011natural}, creating anti-corrosion \cite{cui2018influence} and anti-icing surfaces \cite{huang2022icephobic, yang2013preparation}, water harvesting \cite{wang2022sustainable}, and facilitating oil-water separation \cite{rasouli2021superhydrophobic}. In this context, the present study examines the evaporation dynamics of a two-drop system on a superhydrophobic substrate, owing to the micro-nano surface texturing, under both room and elevated temperature conditions. Additionally, the evaporation behavior of the two-drop system is compared to that of an isolated sessile droplet. When two droplets are in close proximity on the superhydrophobic substrate, their evaporation flux becomes asymmetrical due to the shielding effect \cite{masoud2021evaporation, iqtidar2023drying, schofield2020shielding, chen2022predicting, lee2023vapor}, resulting in slower evaporation of the two-drop system compared to an isolated droplet. Below, we provide a brief review of key findings from existing literature on the evaporation of sessile droplets on surfaces maintained at various temperatures.

Several researchers have explored the evaporation of sessile water droplets and identified three distinct modes, namely constant contact angle (CCA), constant contact radius (CCR), and mixed mode \cite{birdi1989study, picknett1977evaporation, yu2012experimental,hari2022counter}. The foundational studies by \citet{birdi1989study} and \citet{picknett1977evaporation} on the evaporation of sessile water droplets on hydrophilic substrates at room temperature demonstrated a constant evaporation rate. \citet{hu2002evaporation} showed that for droplets with contact angles ($\theta < 40^\circ$), the evaporation rate remained constant over time. \citet{saenz2015evaporation} conducted experiments and numerical simulations, revealing that in the CCA mode, the average interfacial temperature remained relatively constant, whereas in the CCR mode, it showed an increase. \citet{kumar2018combined} examined the evaporation of a sessile water drop on a heated hydrophilic substrate, concluding that evaporative cooling and the temperature dependence of the diffusion coefficient significantly affect the dynamics of the evaporating droplet. On hydrophobic substrates such as Teflon and Polydimethylsiloxane (PDMS), \citet{yu2012experimental} observed that evaporation initially occurred in the constant contact radius (CCR) mode before transitioning to the constant contact angle (CCA) mode and eventually to mixed modes. \citet{bourges1995influence} explored the effect of contact angle on the evaporation of water and decane drops on different substrates. The evaporation dynamics of a sessile droplet on both hydrophilic and hydrophobic substrates have been extensively studied by several researchers \cite{kadhim2019experimental,bhardwaj2018analysis,chatterjee2020evaporation,mahato2023joint}. \citet{gurrala2019evaporation} studied the evaporation of a single binary droplet of the ethanol-water mixture at various compositions and elevated temperatures. The evaporation of multiple droplets at room temperature on hydrophilic and hydrophobic substrates has also been examined by a few researchers \cite{masoud2021evaporation, iqtidar2023drying, wray2020competitive, Malachtari_interaction_2024, shaikeea2016insight, hatte2019universal}. \citet{thokchom2017characterizing} studied the self-assembly and deposition mechanisms of nanoparticles in evaporating inkjet droplets and analyzed the effects of fluid flow, substrate properties, and nanoparticle suspension chemistry on the resulting patterns and structures. \citet{cira2015vapour} investigated the movement of two-component droplets of miscible liquids, such as propylene glycol and water, on clean glass surfaces, driven by evaporation-induced surface tension gradients and free from contact line pinning. Recently, \citet{Hari2024} studied the evaporation of multiple drops on heated hydrophilic and hydrophobic substrates. All these studies focused on the evaporation of sessile droplets on smooth substrates.

The wetting behavior of water droplets on micropillar substrates was investigated by \citet{kumar2020wetting}, who showed that the surface could exhibit hydrophobic or superhydrophobic properties by altering the pitch through radial adjustments. \citet{chuang2014evaporation} examined the evaporation of a water droplet on soft patterned PDMS substrates, finding that the droplet initially evaporated in the CCR mode, transitioned to the CCA mode through contact line jumps, and eventually ended in a mixed evaporation mode. Similarly, \citet{mchale2005analysis} investigated the evaporation of water droplets on superhydrophobic patterned polymer surfaces, observing pinned contact line behavior followed by a stepwise receding of the contact line from pillar to pillar. The evaporation of droplets on carbon nanofiber (CNF) substrates was studied by \citet{gelderblom2011water}, who found that the diffusion-based model of \citet{popov2005evaporative} effectively predicted the evaporation dynamics. \citet{dash2013droplet} observed that the evaporation dynamics of droplets on structured superhydrophobic substrates with minimal contact angle hysteresis followed the CCA mode throughout the evaporation process. While these studies on isolated droplet evaporation were conducted at room temperature, some researchers \cite{dash2014droplet, pan2020transport, erbil2023droplet, yu2024transition} have investigated the evaporation of isolated droplets on superhydrophobic substrates under elevated temperature conditions. With regard to multiple droplets on superhydrophobic substrates, \citet{moradi2020collective} investigated their evaporation dynamics at room temperature. They observed that the evaporation rate on a superhydrophobic surface is slower than on a normal surface for both isolated droplets and droplet collections; however, their study did not explore the theoretical aspects.

As the aforementioned literature review indicates, the evaporation of sessile droplets at both room and elevated substrate temperatures on hydrophilic and hydrophobic surfaces has been extensively studied. The evaporation of multiple droplets at various temperatures on these substrates has also been explored. It is to be noted that the evaporation dynamics of two-drop and multiple-drop systems at elevated temperatures have been previously studied by \citet{hwang2024enhanced} and \citet{Hari2024}, focusing on drops on hydrophilic and mildly hydrophobic substrates (contact angle $\sim 103^\circ$). Experimental work on multiple drops at room temperature on superhydrophobic substrates has been conducted by \citet{moradi2020collective}. However, the evaporation behavior of two drops on superhydrophobic substrates at high temperatures, as well as the theoretical framework for multiple drops evaporating on superhydrophobic substrates at room temperature, remains unexplored, despite its practical significance \cite{roach2008progess, nosonovsky2009superhydrophobic, bhushan2011natural, cui2018influence, huang2022icephobic, yang2013preparation, wang2022sustainable, rasouli2021superhydrophobic}. In the present study, we investigate the evaporation dynamics of a two-droplet system placed in close proximity on micro-nano textured superhydrophobic substrates at both room and elevated temperatures, a topic that, to the best of our knowledge, has not been explored previously. Moreover, we examine the lifetimes of the droplets in the two-drop system and compare them with the lifetimes of isolated droplets under the same conditions. Using shadowgraphy, we capture the size profile of the droplets and analyze the temporal evolution of the contact diameter, contact angle, and volume for both the two-drop and isolated droplet systems. Typically, diffusion-based models predict the evaporation dynamics at room temperature, irrespective of the different modes like constant contact angle (CCA), constant contact radius (CCR), and mixed-mode evaporation \cite{hwang2017droplet}. Therefore, we employ a diffusion-based theoretical model to compare the evaporation dynamics of single and multiple droplets on superhydrophobic substrates with the experimental data, demonstrating satisfactory agreement. The details of the experimental methodology are provided in the following section.

\section{Experimental Methodology}
\label{sec:expt}

\begin{figure}[h]
\centering
\includegraphics[width=0.9\textwidth]{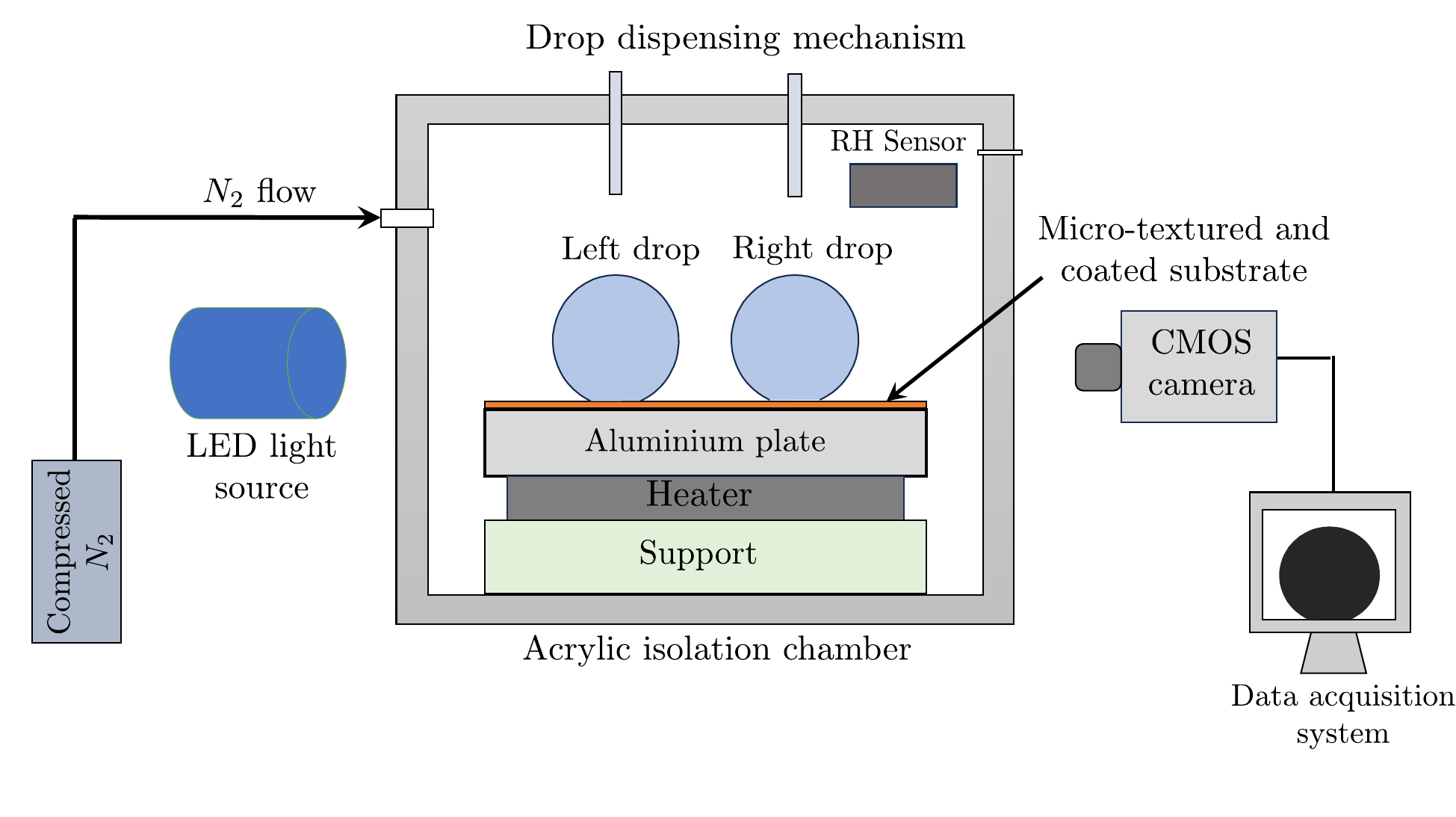}
\caption{Schematic diagram of the experimental setup. It consists of a droplet dispensing mechanism connected to two syringe pumps, enabling synchronized droplet deposition onto a temperature-controlled substrate managed by a heating system hosted within an acrylic chamber. Humidity control is achieved by supplying $N_2$ to the chamber. Imaging is performed using shadowgraphy, which incorporates a CMOS camera paired with an LED light source.}
\label{fig:fig1}
\end{figure}

We investigate the evaporation dynamics of sessile water droplets on a micro-nano textured superhydrophobic aluminum substrate using the shadowgraphy imaging technique. The study focuses on the evaporation behavior of a system of two droplets placed side-by-side on a superhydrophobic substrate maintained at both room and elevated temperatures. The evaporation dynamics of the two-drop system are compared with those of an isolated droplet. A schematic of the experimental setup is shown in Figure \ref{fig:fig1}. The experimental setup consists of an acrylic isolation chamber designed to maintain a constant relative humidity. Nitrogen gas ($N_{2}$) from a compressed cylinder is supplied to the chamber, with the flow rate regulated by a Swagelok mass flow control valve (Model: SS-$4$MG-MH $0003978175$) and a shut-off valve (Model: SS-$4$P$4$T $0004034885$). The $N_{2}$ gas is continuously purged into the chamber and exits through a small opening on the opposite side, ensuring a controlled inflow that maintains the relative humidity at $16 \pm 2\%$. The side view of the evaporating droplets is captured by a Complementary Metal Oxide Semiconductor (CMOS) camera (Model: Thorlabs Zelux CS165CU1/M) with a resolution of $1440 \times 1080$ pixels at 35 frames per second (fps). The camera is equipped with a Navitar zoom lens and is illuminated by a Light Emitting Diode (LED) light assembly. 

In our experiment, water droplets are dispensed using a 34-gauge needle connected to two syringe pumps, with a dispensing rate of 4.0 $\mu$l/min. The initial volume of the droplets generated using this mechanism is $V_0 = 4.0 \pm 0.3$ $\mu$l when deposited onto the substrate. In the two-drop configuration, the interval between dispensing the droplets is limited to 5 seconds, which is significantly shorter than the total evaporation time for each droplet. For an isolated droplet, the average initial diameter $(d_{0})$ on the substrate at room temperature ($27^\circ$C) is $2.088 \pm 0.015$ mm, whereas, when the substrate is heated ($50^\circ$C), $d_{0}$ is $2.047 \pm 0.013$ mm. In the two-drop configuration, at room temperature, $d_{0} = 2.096 \pm 0.012$ mm, while at $50^\circ$C, $d_{0} = 2.068 \pm 0.017$ mm. The error bars are obtained by performing three repetitions for each set of parameters. Since the droplets are dispensed simultaneously and the contact area on a superhydrophobic substrate is minimal compared to hydrophobic and hydrophilic surfaces, the variation in initial droplet diameter between room temperature and heated substrates is expected to be insignificant. 

The evaporation dynamics of the droplets are recorded using a CMOS camera. The substrate temperature is controlled using a heater (PTC Heaters Plate 24V 70 Degrees) powered by a DC supply. The temperature regulation is achieved through precise control of the heater, while a PT1000 RTD sensor monitors the substrate temperature. Additionally, the chamber is equipped with a humidity sensor (Model: Rotronic HC2A-S) to measure the relative humidity, ensuring consistent environmental conditions throughout the experiments. 

\begin{figure}[h]
\centering
\includegraphics[width=0.6\textwidth]{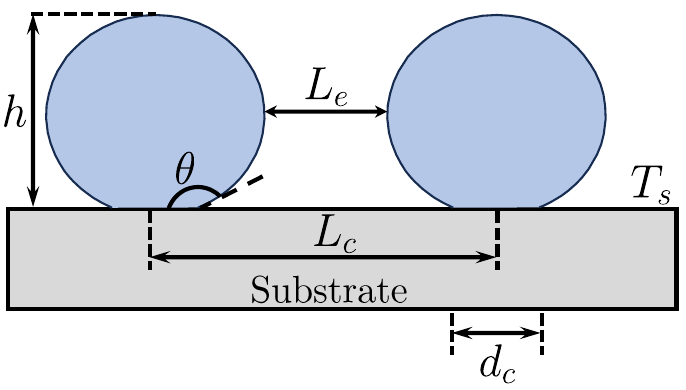}
\caption{Schematic representation of a pair of drops on a superhydrophobic substrate. In this configuration, $d_{c}$ denotes the wetting diameter of the droplets, $L_{e}$ represents the distance between the edges of the two drops ($L_e \leq 0.37\ \text{mm}$), and $L_{c}$ is the distance between the centers of the drops. Here, $\theta$ and $h$ denote the contact angle and height of the drops, respectively and $T_s$ represents the temperature at the op surface of the substrate..}
\label{fig:fig2}
\end{figure}

In order to prepare the substrates, first, Acetone, isopropyl alcohol, and de-ionised water were used to ultrasonically clean the $6.5 \times 3 \times 0.3$ cm aluminium substrates (Aluminium 6062) for $10$ minutes each. The sheets were then etched dissociation-selectively in a $1$ M FeCl$_{3}$ solution at $25^\circ$C for $25$ minutes after being treated in a NaOH solution for $10$ minutes \cite{maitra2014hierarchically,sharma2017growth}. Ultrasonic cleaning with isopropyl alcohol was performed every $2.5$ minutes during the etching process to develop a microstructure on the aluminium substrate. Subsequently, an overlying nanostructure was created using the boehmitization method. To achieve superhydrophobicity, the samples were immersed in a $1.43$ M solution of trichloro-$1$H,$1$H,$2$H,$2$H-perfluorodecylsilane (FDTS, Sigma-Aldrich) in $n$-hexane for $2$ hours \cite{maitra2014hierarchically,sharma2017growth}. Finally, the samples were baked at $120^\circ$C for $45$ minutes. The micro-nano textured superhydrophobic aluminium substrate exhibited a contact angle ranging from $155^\circ$ to $165^\circ$, with a contact angle hysteresis of approximately $6^\circ$. The evaporation dynamics of both isolated droplets and two-droplet systems were studied on the micro-nano structured aluminium substrate at two substrate temperatures: $T_s = 27 \pm 1^\circ$C and $T_s = 50 \pm 1^\circ$C, with the relative humidity maintained at $RH = 16 \pm 2$\%.

\begin{figure}
\centering
 \hspace{0.1cm}  {\large (a)} \hspace{6.8cm} {\large (b)} \\
 \includegraphics[width=0.441\textwidth]{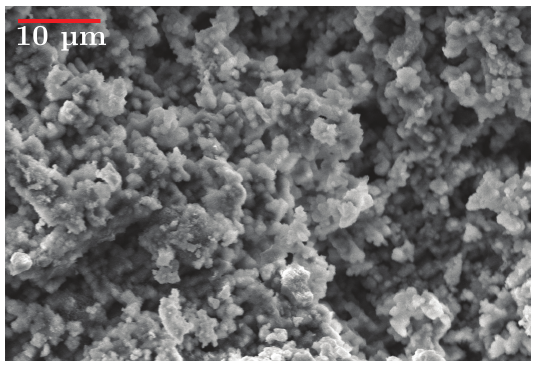} \hspace{2mm} \includegraphics[width=0.425\textwidth]{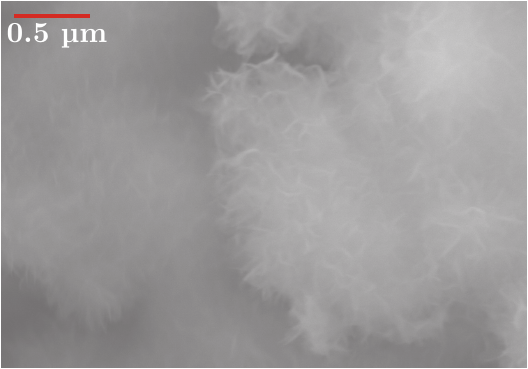}
 \caption{The SEM images of the micro and nano textured aluminum substrates are presented at two magnifications: (a) low magnification, showing the micro-pores, and (b) high magnification, offering a closer view that reveals the presence of bohemite nanostructures.}
\label{fig:fig3}
\end{figure}

A schematic representation of a pair of drops placed in a closed vicinity on a superhydrophobic substrate representing various physical parameters is depicted in Figure \ref{fig:fig2}. Here, $d_{c}$  denotes the dimater of the contact line, i.e. the base diameter of the droplets, $L_{e}$ represents shortest distance between the surfaces of the two drops, which is maintained at $L_e \leq 0.37\ \text{mm}$, and $L_{c}$ is the distance between the centers of the drops. The contact angle is denoted by $\theta$, the height of the drops by $h$, and the temperature at the top surface of the substrate by $T_s$. The substrate is characterized using Scanning Electron Microscopy (SEM), and the surface images at different magnification levels are shown in Figure \ref{fig:fig3}(a) and (b). The side-view profiles of the droplets were extracted by performing post-processing on the side-view images captured by the CMOS camera, using a custom program developed within the \textsc{Matlab}$^{\circledR}$ framework. This process involved applying median filtering to remove random noise and unsharp masking to enhance image sharpness and gradients. The filtered image was then converted to binary format using an appropriate threshold to differentiate the background from the droplet boundary. Holes within the droplet boundary were filled, and the droplet reflection was eliminated. A \textsc{Matlab}$^{\circledR}$ function was utilized to trace the droplet contour, from which various droplet parameters were extracted. This post-processing procedure is similar to the method described by \citet{gurrala2019evaporation,Hari2024}.

\section{Results and Discussion}\label{sec:results}

\begin{figure}[h]
\centering
\includegraphics[width=0.5\textwidth]{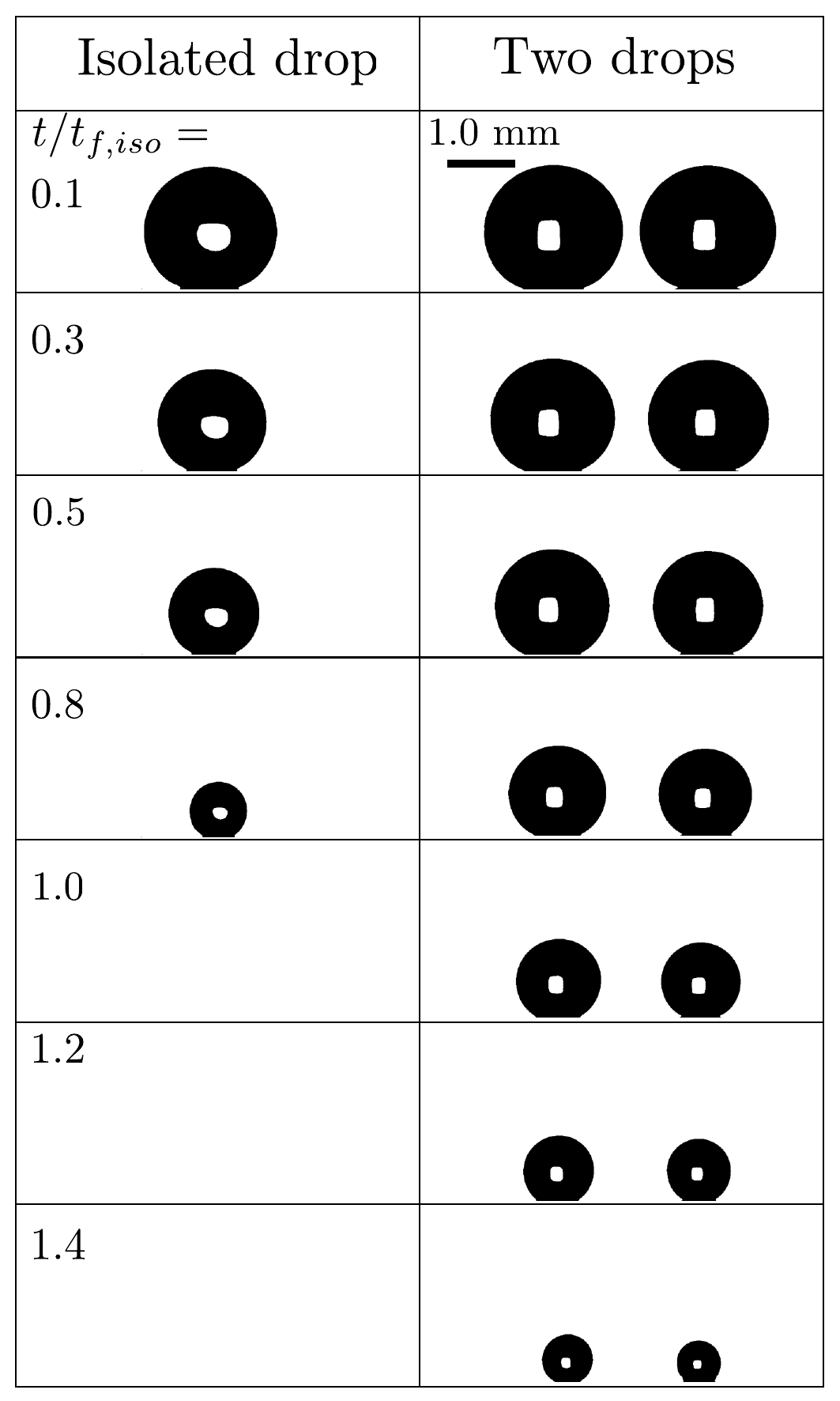}
\caption{Temporal evolution of the droplet shapes undergoing evaporation in both the isolated and two-drop configurations at $T_s = 27^\circ$C and $RH = 0.16$. Here, $t_{f,iso} = 1280$ s represents the total evaporation time of the isolated drop.} 
\label{fig:fig4}
\end{figure}

\begin{figure}[h]
\centering
\includegraphics[width=0.5\textwidth]{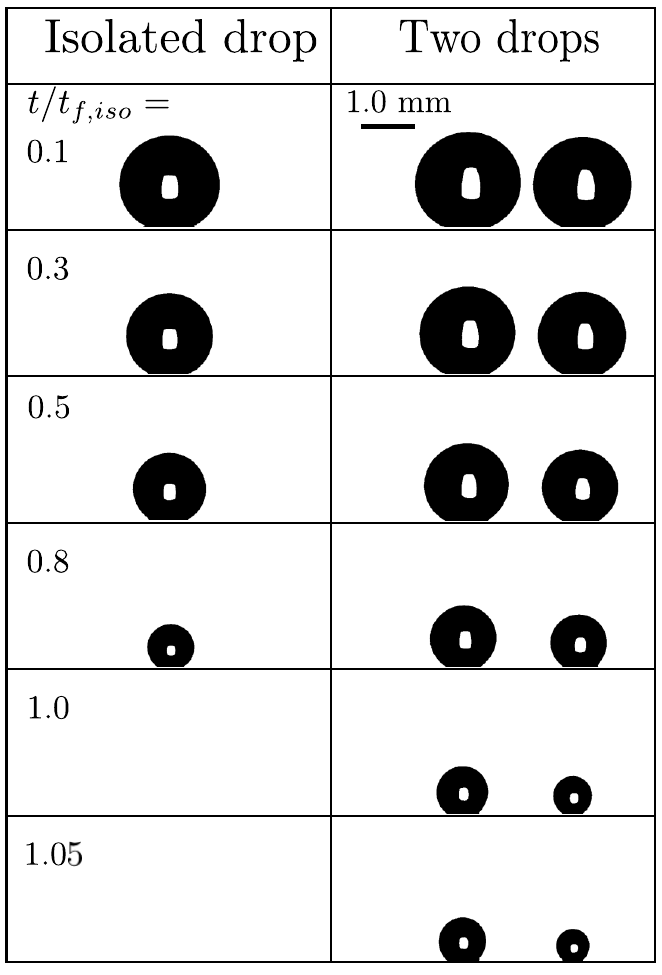}
\caption{Temporal evolution of the droplet shapes undergoing evaporation in both the isolated and two-drop configurations at $T_s = 50^\circ$C and $RH = 0.16$. Here, $t_{f,iso} = 498$ s represents the total evaporation time of the isolated drop.} 
\label{fig:fig5}
\end{figure}

We begin by presenting our results with an analysis of the lifetime and morphology of droplets in both isolated and two-drop systems on a micro-nano textured superhydrophobic substrate maintained at a) $T_s = 27^\circ$C and b) $T_s = 50^\circ$C. In our two-drop experiments, the droplets are placed in close proximity, with $L_c/d_c < 2.87$ mm and $L_e \leq 0.37$ mm, as shown in Figure \ref{fig:fig2}. We perform at least three repetitions for each set of parameters while maintaining a constant relative humidity of $RH = 16$\% inside the chamber. Table \ref{table:T1} presents the evaporation lifetimes of the isolated droplet ($t_{f,iso}$) and the average lifetime of both the drops in the two-drop configuration at both substrate temperatures. We found that at room temperature ($T_s = 27^\circ$C), the isolated droplet evaporates in $t_{f,iso} = 1269 \pm 18$ s, whereas in the two-drop system, the average lifetime of both droplets is $1.6$ times that of the isolated droplet. At a higher substrate temperature ($T_s = 50^\circ$C), the isolated droplet evaporates in $t_{f,iso} = 494 \pm 24 ~ {\rm s}$ , and the average lifetime of both droplets in the two-drop system is $1.2$ times the evaporation time of the isolated droplet. To analyze the evaporation dynamics, we examine the temporal evolution of the droplet morphology for both the isolated and two-drop systems at $T_s = 27^\circ$C (Figure \ref{fig:fig4}) and $T_s = 50^\circ$C (Figure \ref{fig:fig5}). We observe that at $T_s = 27^\circ$C, $L_{c} \approx 2.46$ mm and $L_{e} \approx 0.22$ mm, while at $T_s = 50^\circ$C, $L_{c} \approx 2.27$ mm and $L_{e} \approx 0.12$ mm. In Figures \ref{fig:fig4} and \ref{fig:fig5}, it can be seen that the volume of the isolated droplet is significantly smaller compared to the droplets in the two-drop system at all the instants. This difference is attributed to the vapor shielding effect (see, Refs. \cite{schofield2020shielding,wray2020competitive,wilson2023evaporation,chen2022predicting,hwang2024enhanced}), which substantially increases the vapor concentration between the two droplets. This results in an asymmetric evaporation flux between the inner and outer edges of the droplets in the two-drop system, leading to a slower overall evaporation rate for the two-drop system. Additionally, it can be observed that when the substrate temperature is increased to $T_s = 50^\circ$C while keeping the relative humidity constant, the vapor shielding effect continues to play a role, but it is less pronounced compared to the case at room temperature ($T_s = 27^\circ$C). This is evident in Figure \ref{fig:fig5}, where, at $t/t_{f,iso} = 0.3$, $0.5$, and $0.8$, the size of the isolated droplet remains smaller than the drops in the two-drop system, though the difference is less significant than at room temperature. This behavior is partly due to enhanced natural convection, which increases evaporation rates even for isolated water droplets, as reported by \citet{josyula2018evaporation,pan2020transport}. At elevated temperatures, the two-drop system transitions more rapidly to isolated-drop behavior compared to room temperature. This aspect will be further discussed at the end of this section. Natural convection, driven by buoyant airflow along the substrate, reduces the local relative humidity and thereby promotes evaporation. Notably, for organic liquids~\cite{shahidzadeh2006evaporating} and more volatile fluids, convective transport plays a significant role in evaporation even at room temperature~\cite{kelly2011evaporation,carle2013experimental}. Throughout the evaporation process, the superhydrophobic nature of the drop (high contact angle) remains largely unchanged until the end of evaporation, as seen in Figures \ref{fig:fig4} and \ref{fig:fig5}. 

\begin{table}
\centering
\caption{Lifetimes of the droplets (in seconds) for the isolated drop and the average lifetime of the drops in the two-drop system on a substrate maintained at $T_s = 27^\circ$C and $T_s = 50^\circ$C.} 
\label{table:T1}
\hspace{0.0cm}   \\
\begin{tabular}{|c|c|c|c|c|c}\hline
$T_s$  & Lifetime of an isolated drop& Lifetime in two-drop configuration  \\ \cline{1-3}
27$^\circ$C & $1269\pm18$ s& $2050\pm36$ s \\ \cline{1-3}
50$^\circ$C & $494\pm24$ s & $593\pm41$ s \\ 
\hline
\end{tabular}
\end{table}

\begin{figure}[h]
\centering
\hspace{0.75cm}{\large (a)} \hspace{6.75cm}  {\large (b)} \\
\includegraphics[width=0.45\textwidth]{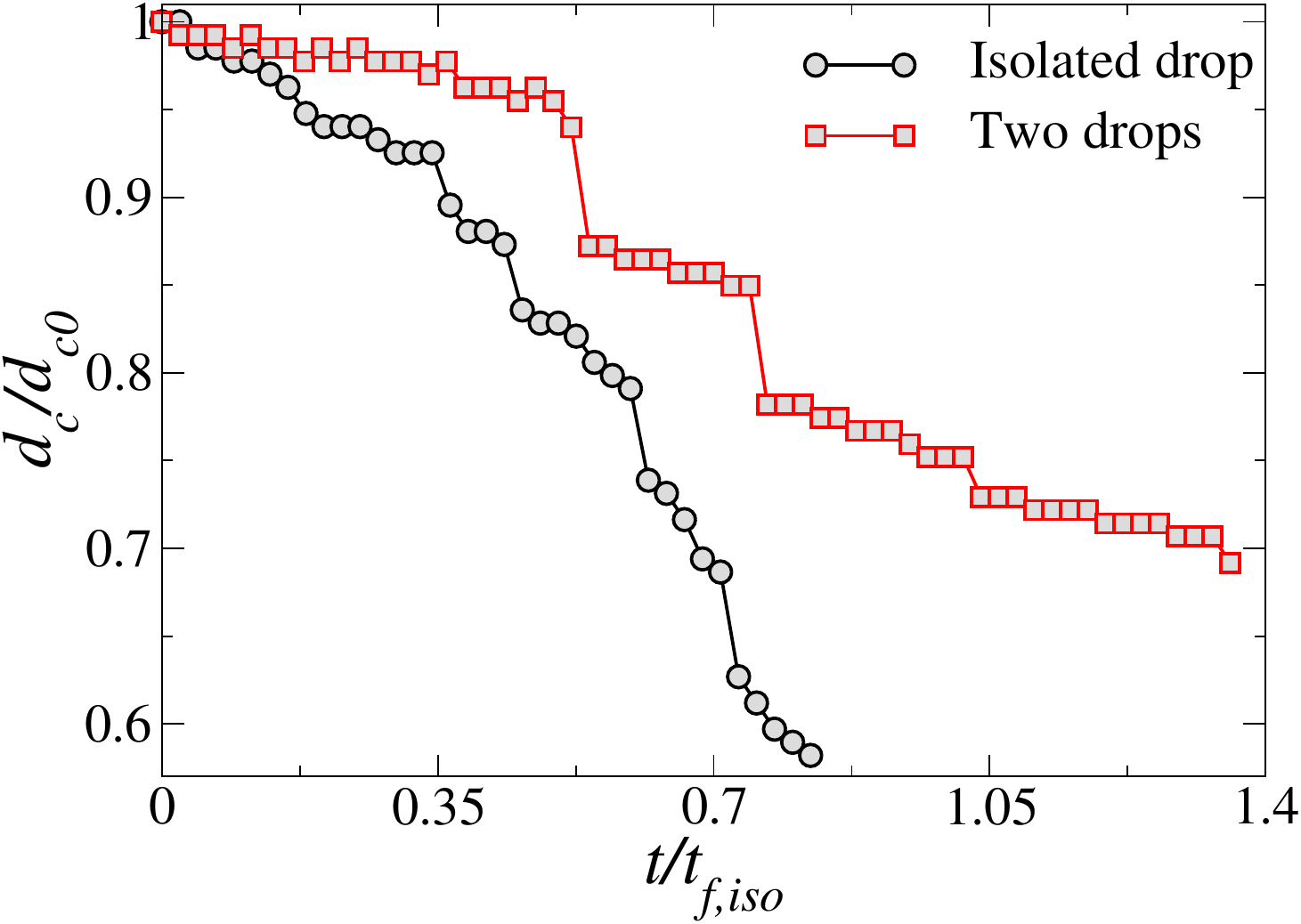} 
\hspace{0mm}
\includegraphics[width=0.45\textwidth]{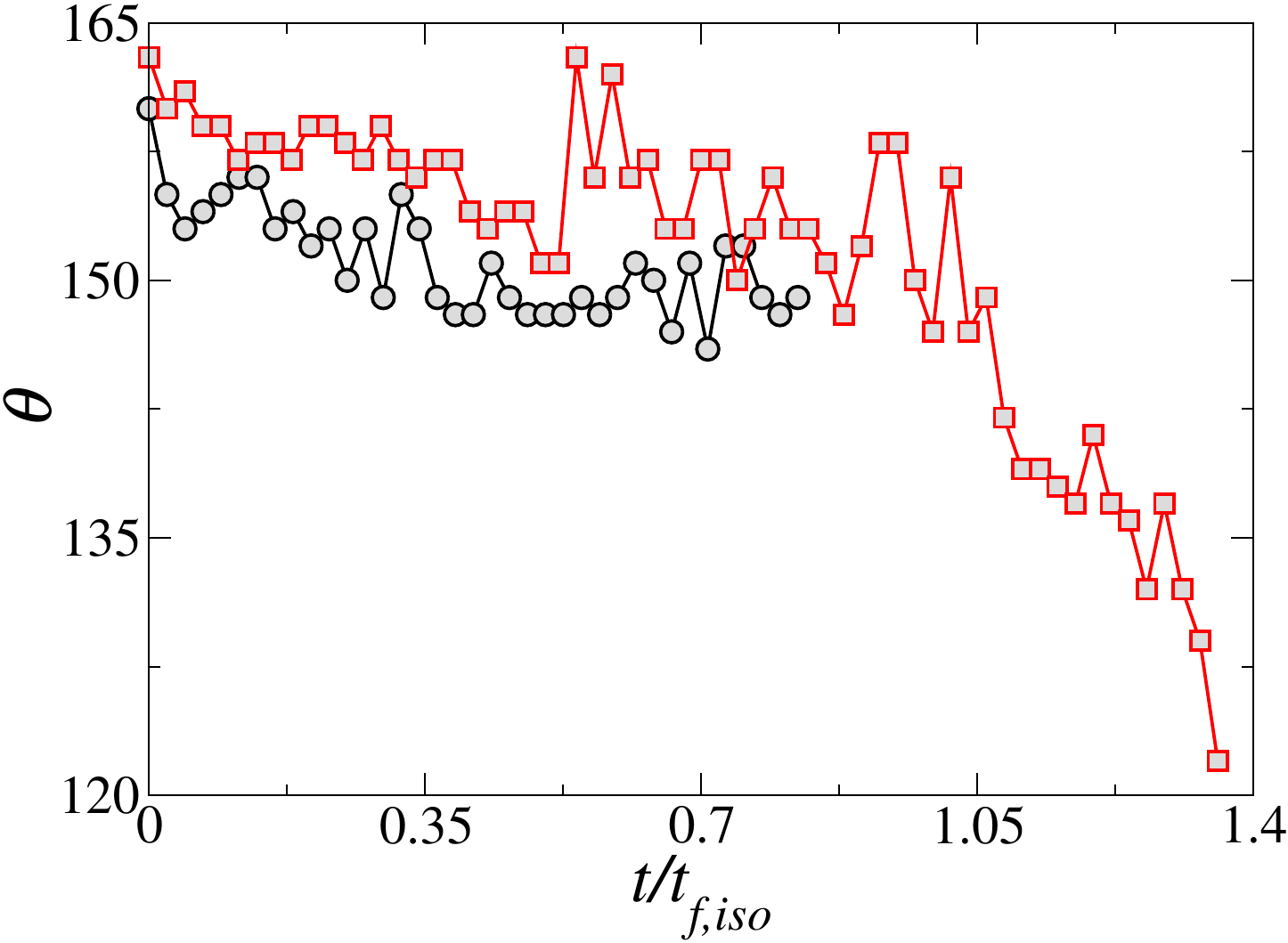}\\
\caption{Temporal variation of (a) normalized wetting diameter ($d_c/d_{c0}$) and (b) contact angle ($\theta$) for the isolated drop and the left drop of the two-drop configuration at $T_s = 27^\circ$C. The chamber is maintained at $RH = 0.16$. Here, $t_{f,iso} = 1280$ s represents the total evaporation time of the isolated drop.}
\label{fig:fig6}
\end{figure}

To gain further insight into the evaporation dynamics, we analyze the normalized wetting diameter ($d_c/d_{c0}$), contact angle ($\theta$), and normalized volume ($V/V_0$) of both the isolated droplet and the two-drop system using side-view imaging data from the CMOS camera. Here, $d_c$ and $d_{c0}$ represent the instantaneous and initial wetting diameters of the droplet, respectively. Similarly, $V$ and $V_0$ represent the instantaneous and initial volumes of the droplet. It is important to note that the droplet geometric parameters can be accurately measured up to 80\% of their total lifetime due to the maximum resolution of the recorded images. Therefore, we present the data only up to that point. Figures \ref{fig:fig6}(a) and (b) show the temporal variation of the normalized droplet wetting diameter ($d_{c}/d_{c0}$) and contact angle ($\theta$) for both the isolated droplet and the two-drop system at room temperature ($T_{s} = 27^\circ$C). In this figure, the data of one repetition is presented in the main manuscript, while the other two repetitions for both the isolated droplet and two-drop system at $T_{s} = 27^\circ$C are provided in supplementary Figure S1. As shown in Figures \ref{fig:fig6}(a) and (b), the isolated droplet at room temperature primarily follows a constant contact angle (CCA) mode of evaporation, with a few instances of stick-slip behavior. For the two-drop system, the contact angle remains nearly constant until $t/t_{f,iso} = 1$, after which it decreases significantly, as shown in Figure \ref{fig:fig6}(b). The two-drop system at room temperature predominantly exhibits the CCA mode for most of its lifetime, with intermittent phases of mixed-mode evaporation and occasional stick-slip behavior, as seen in Figure \ref{fig:fig6}(a) and (b). It is to be noted that several factors, such as the random distribution of nano and micropores (shown in Figure \ref{fig:fig3}), the local roughness of the substrate, and the asymmetry in the evaporation flux can influence the stick-slip behavior observed in the present study. A demonstration of this behavior for an isolated droplet at $T_s = 27^\circ$C is shown in Figure S2. 

\begin{figure}[h]
\centering
\hspace{0.75cm}{\large (a)}   \hspace{6.75cm}  {\large (b)} \\
\includegraphics[width=0.45\textwidth]{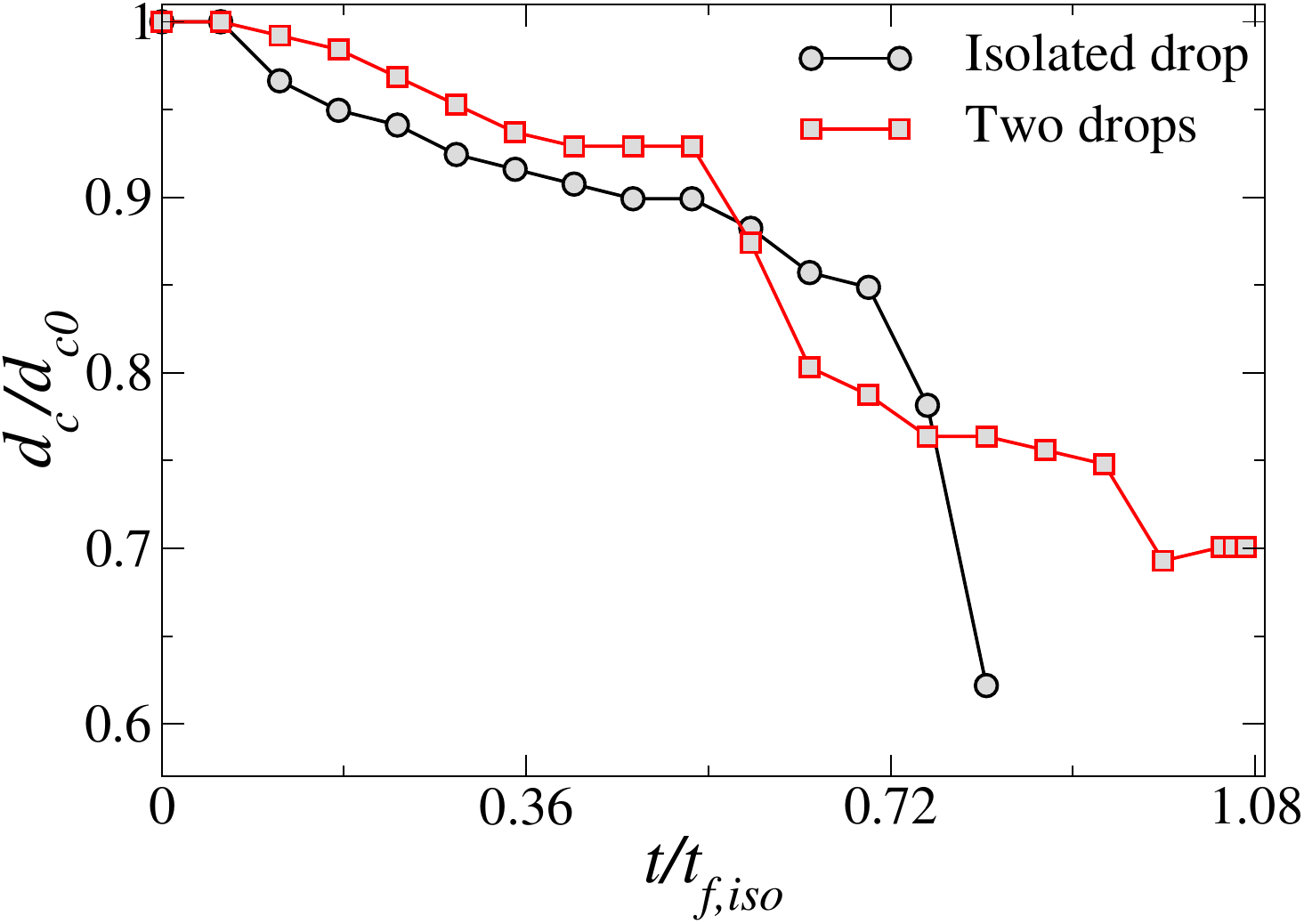}
\hspace{0mm}
\includegraphics[width=0.45\textwidth]{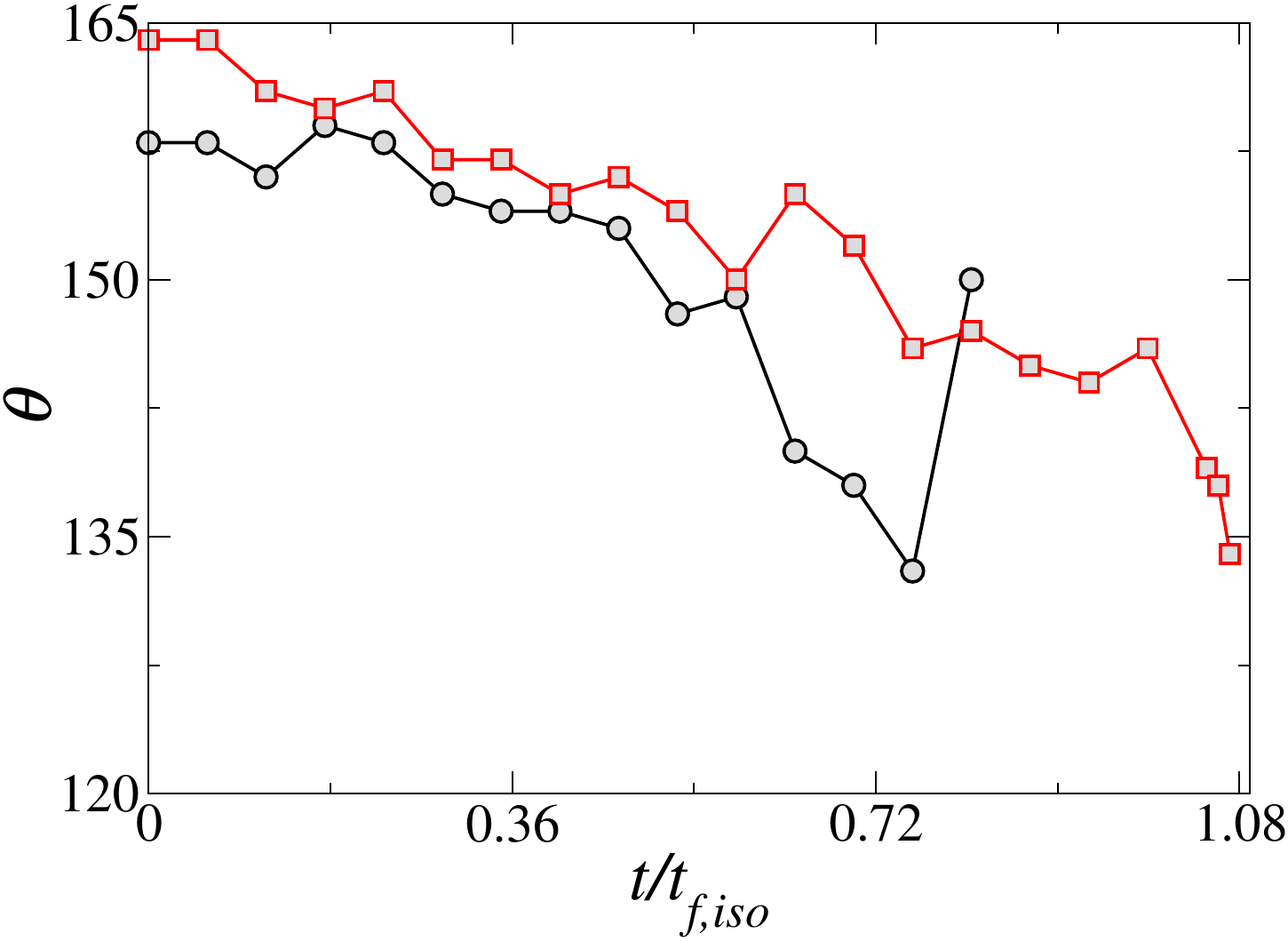}\\
\caption{Temporal variation of (a) normalized wetting diameter ($d_c/d_{c0}$) and (b) contact angle ($\theta$) for the isolated drop and the left drop of the two-drop configuration at $T_s = 50^\circ$C. The chamber is maintained at $RH = 0.16$. Here, $t_{f,iso} = 498$ s represents the total evaporation time of the isolated drop.}
\label{fig:fig7}
\end{figure}

At high substrate temperature ($T_s = 50^\circ$C), as shown in Figure \ref{fig:fig7}, the isolated drop maintains a nearly constant contact angle for half of its lifetime, after which it transitions to a mixed mode of evaporation with some instances of stick-slip behavior. For the two-drop system, the evaporation dynamics follow a mixed mode throughout most of its lifetime, with occasional stick-slip events, as observed in Figure \ref{fig:fig7}(a) and \ref{fig:fig7}(b). The dynamics of a pair of drops on a hydrophobic substrate with a contact angle of $115^\circ$ was investigated by \citet{shaikeea2016insight} at room temperature. Their findings revealed that the drops exhibited the constant contact radius (CCR) mode during the initial and final stages of evaporation, while the mixed mode dominated the intermediate stages of the evaporation cycle. \citet{Hari2024} explored the evaporation behavior of multiple drops on a hydrophobic substrate with a contact angle of $103^\circ$ and observed that the drops initially followed a CCR mode, transitioning to a constant contact angle (CCA) mode at both room temperature and elevated substrate temperatures. In the present study, we observe that, at room temperature, the two-drop system primarily displayed a CCA mode, interspersed with phases of mixed-mode evaporation and intermittent stick-slip behaviour. At high temperatures, the evaporation dynamics of the two-drop system were predominantly governed by the mixed mode throughout most of its lifetime, with occasional occurrences of stick-slip behavior. The results obtained from different repetitions at the high substrate temperature ($T_s = 50^\circ$C) are provided in supplementary Figure S3 for both the isolated and two-drop systems. Furthermore, in Figure S4 of the supplementary information, we investigate the differences in contact line dynamics between isolated and two-drop systems at different substrate temperatures ($T_s = 27^\circ$C and $50^\circ$C). The temporal variations of the normalized wetting diameter ($d_c/d_{c0}$) and contact angle ($\theta$) for the isolated drop are presented in Figure S4(a) and Figure S4(b), respectively, while the corresponding results for the two-drop system are shown in Figure S4(c) and Figure S4(d). We observe that, for the isolated drop, the normalized wetting diameter ($d_c/d_{c0}$) remains nearly identical up to $t/t_{f,\mathrm{iso}} = 0.4$. Beyond this point, the drop on the heated substrate exhibits a larger normalized wetting diameter compared to that on the substrate maintained at room temperature (Figure S4(a)). As shown in Figure S4(b), the contact angle ($\theta$) remains similar at both temperatures up to $t/t_{f,\mathrm{iso}} = 0.6$, after which the angle on the heated substrate decreases and then increases, indicating a stick-slip behaviour during the evaporation process. For the two-drop system (Figure S4(c)), the normalized wetting diameter remains similar at both temperatures up to $t/t_{f,\mathrm{iso}} = 0.5$. Thereafter, the wetting diameter on the heated substrate decreases rapidly, due to the significantly shorter evaporation time at the higher temperature. The contact angle variation for the two-drop system (Figure S4(d)) shows minimal difference between the two substrate temperatures. Notably, when the evaporation time for the two-drop system is normalized by its total lifetime (rather than that of the isolated drop), the temporal variation of the normalized wetting diameter ($d_c/d_{c0}$) becomes nearly identical at $T_s = 27^\circ$C and $T_s = 50^\circ$C throughout the droplet lifetime.

\begin{figure}
\centering
 \hspace{0.5cm}  {\large (a)} \hspace{7.1cm} {\large (b)} \\
 \includegraphics[width=0.45\textwidth]{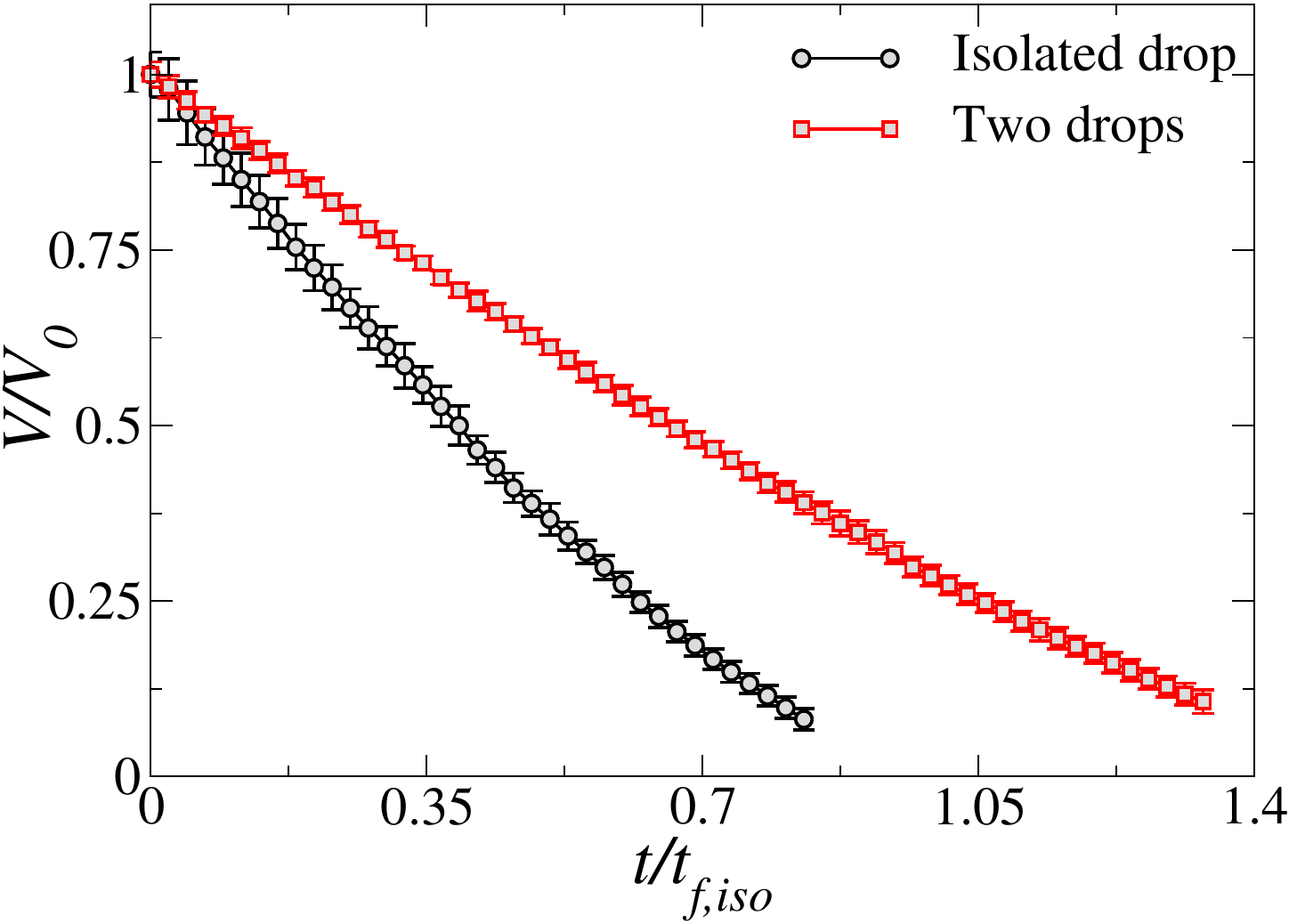} \hspace{2mm} \includegraphics[width=0.45\textwidth]{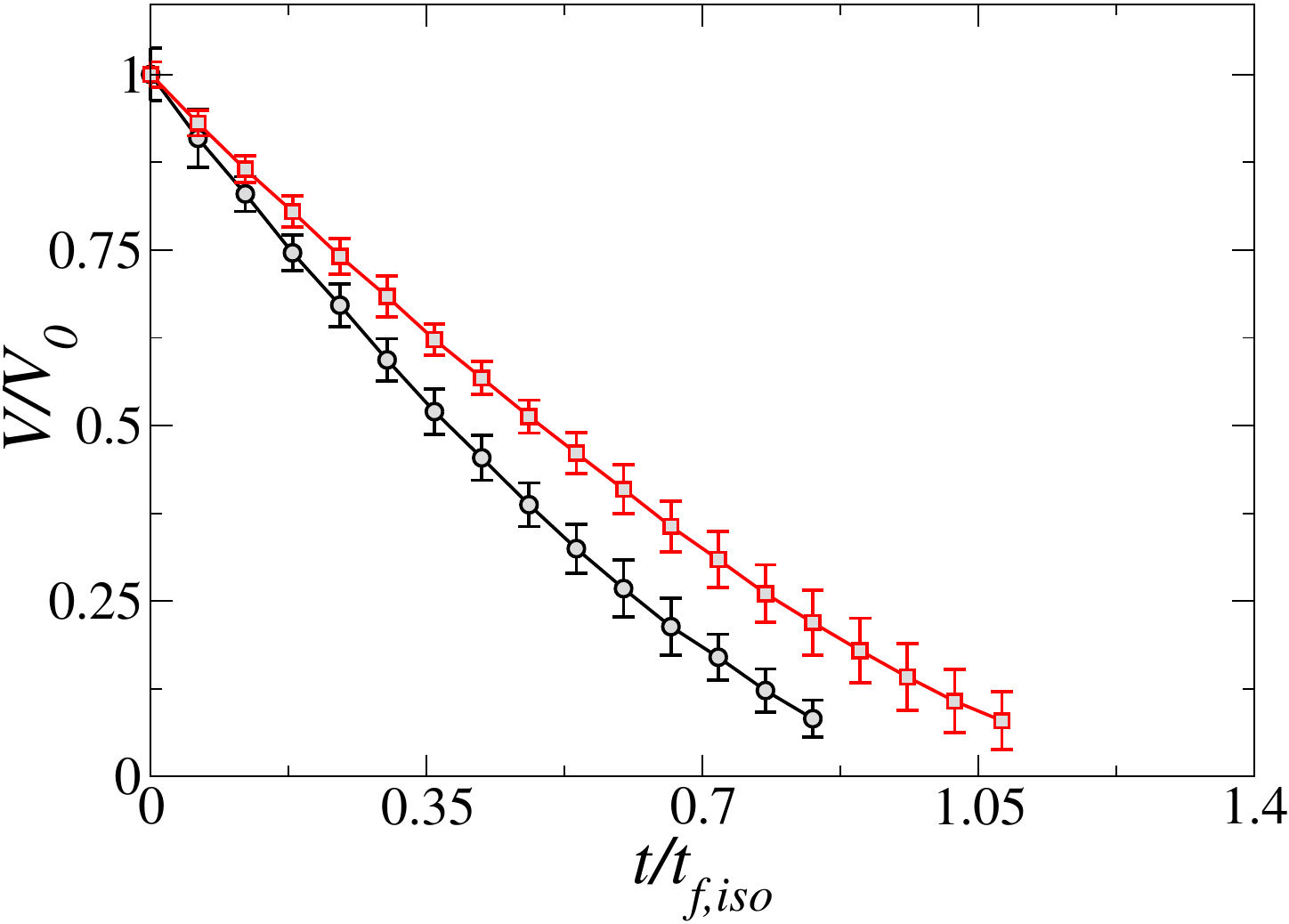}
\caption{Variation of the normalized droplet volume ($V/V_0$) of the left drop with $t/t_{f,iso}$ for the isolated drop and two-drop configurations at (a) $T_s = 27^\circ$C and (b) $T_s = 50^\circ$C. The chamber is maintained at a relative humidity of $RH = 0.16$. Here, $V_0$ represents the initial volume of the drop in the isolated drop case and the initial volume of the left drop in the two-drop configuration.}
\label{fig:fig8}
\end{figure}

\begin{figure}
\centering
 \hspace{0.5cm}  {\large (a)} \hspace{7.1cm} {\large (b)} \\
 \includegraphics[width=0.45\textwidth]{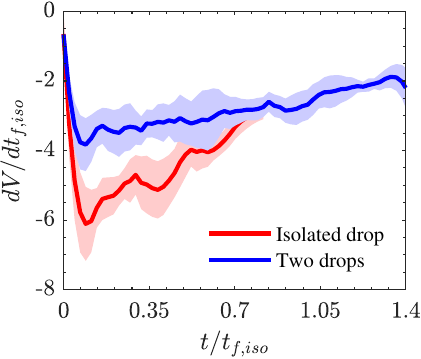} \hspace{2mm} \includegraphics[width=0.45\textwidth]{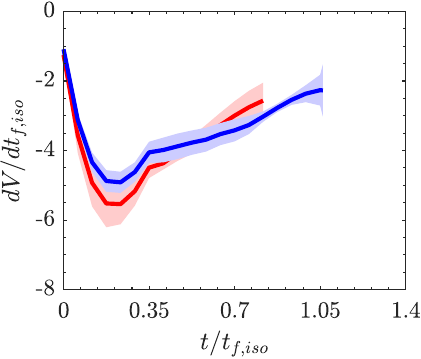}
\caption{The variation of the rate of change of the volume of the isolated drop and the left drop of the two-drop system with $t/t_{f,iso}$ at (a) $T_s = 27^\circ$C and (b) $T_s = 50^\circ$C. The chamber is maintained at a relative humidity of $RH = 0.16$. The shaded regions represent the uncertainty across multiple experimental measurements.}
\label{fig:fig9}
\end{figure}

Figure \ref{fig:fig8}(a) shows the temporal variation of the normalized volume ($V/V_{0}$) for the isolated drop and the two-drop system. The isolated drop shows a steeper monotonic decrease in normalized volume ($V/V_{0}$) compared to the two-drop system, which takes around $60\%$ more time to evaporate than the isolated drop at room temperature. As the substrate temperature increases to $T_s = 50^\circ$C, the normalized volume ($V/V_0$) of both the isolated drop and the two-drop system decreases monotonically, with only a $20\%$ increase in the lifetime of the two-drop system, as shown in Figure \ref{fig:fig8}(b). In Figure \ref{fig:fig9}(a) and (b), we show that, at later stages of evaporation, the two-drop system begins to behave like two isolated drops. This is evident as the rate of change of volume ($dV/dt_{f,\text{iso}}$) with scaled time ($t/t_{\text{iso}}$) for the left drop in the two-drop system converges to the rate of change of volume ($dV/dt_{f,\text{iso}}$) with scaled time ($t/t_{\text{iso}}$) for the isolated drop at later times, for both $T_s = 27^\circ$C and $T_s = 50^\circ$C. Figures~\ref{fig:fig9}(a) and \ref{fig:fig9}(b) illustrate that, for the two-drop system at room temperature ($T_s = 27^\circ$C), the drops begin to behave like isolated drops after $t/t_{f,\text{iso}} = 0.7$. In contrast, at the higher substrate temperature ($T_s = 50^\circ$C), this transition occurs earlier, at approximately $t/t_{f,\text{iso}} \approx 0.4$. This earlier transition can be attributed to enhanced natural convection effects at elevated temperatures. Furthermore, the difference in the rate of volume change ($dV/dt_{f,\text{iso}}$) with respect to the scaled time ($t/t_{f,\text{iso}}$) between the two-drop system and the isolated drop is smaller at higher substrate temperatures, as shown in Figures~\ref{fig:fig9}(a) and \ref{fig:fig9}(b). This suggests that the influence of the vapor shielding effect is reduced at elevated substrate temperatures.

\subsection{Theoretical modeling}
\begin{figure}
\centering
 \hspace{0.5cm}  {\large (a)} \hspace{7.1cm} {\large (b)} \\
 \includegraphics[width=0.45\textwidth]{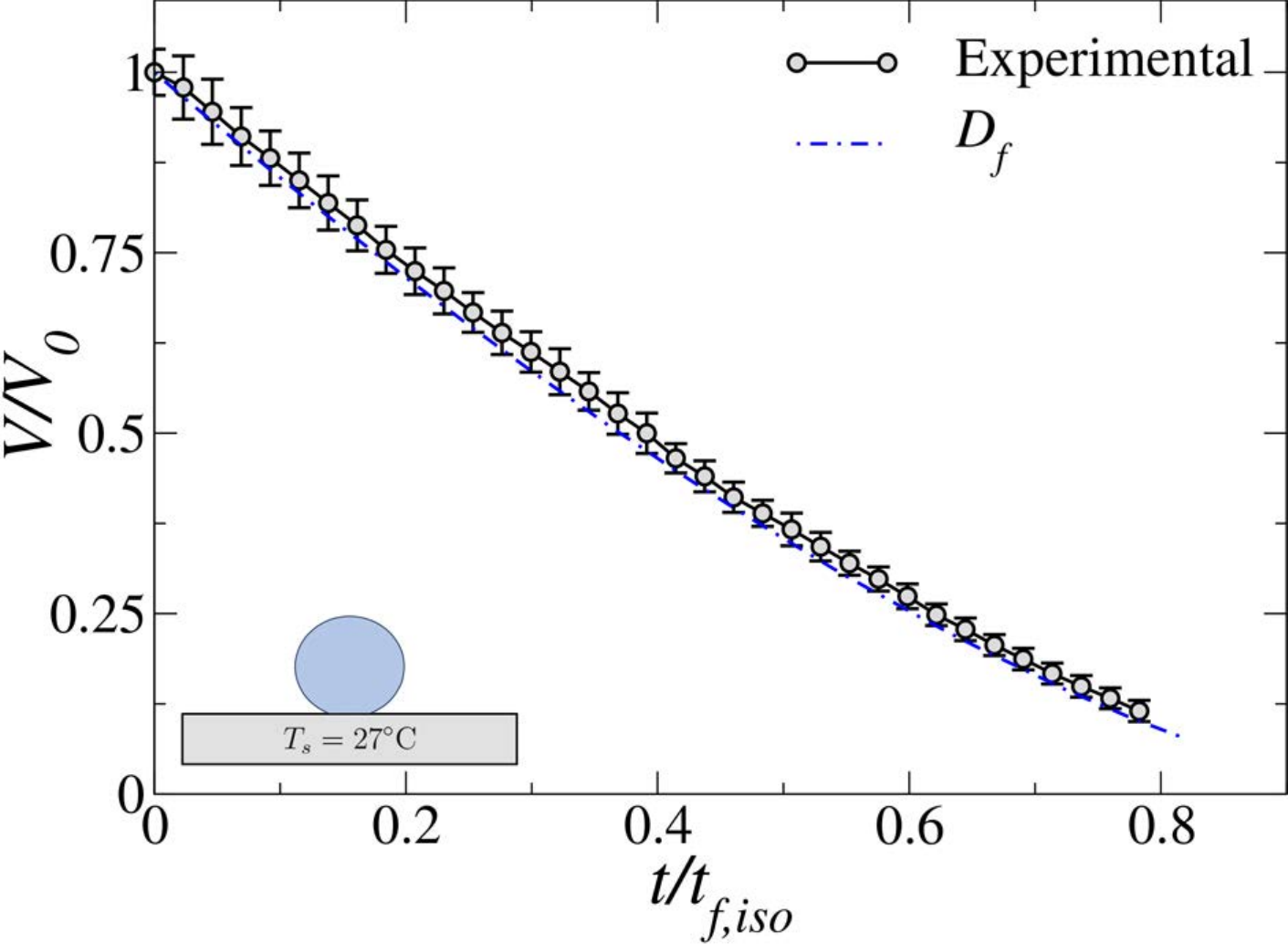} \hspace{2mm} \includegraphics[width=0.45\textwidth]{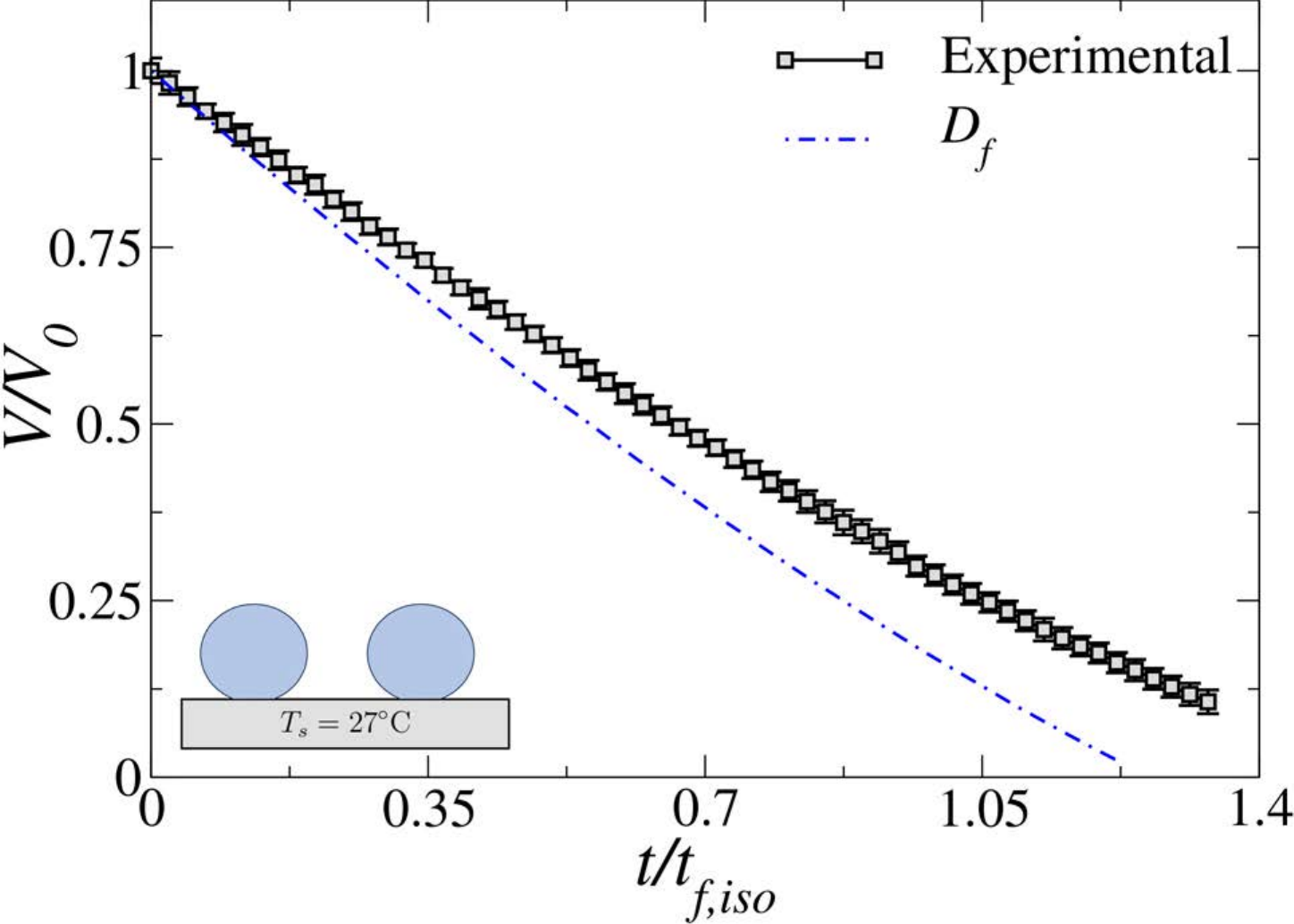} \\
  \hspace{0.5 cm}  {\large (c)} \hspace{7.1cm} {\large (d)} \\
 \includegraphics[width=0.46\textwidth]{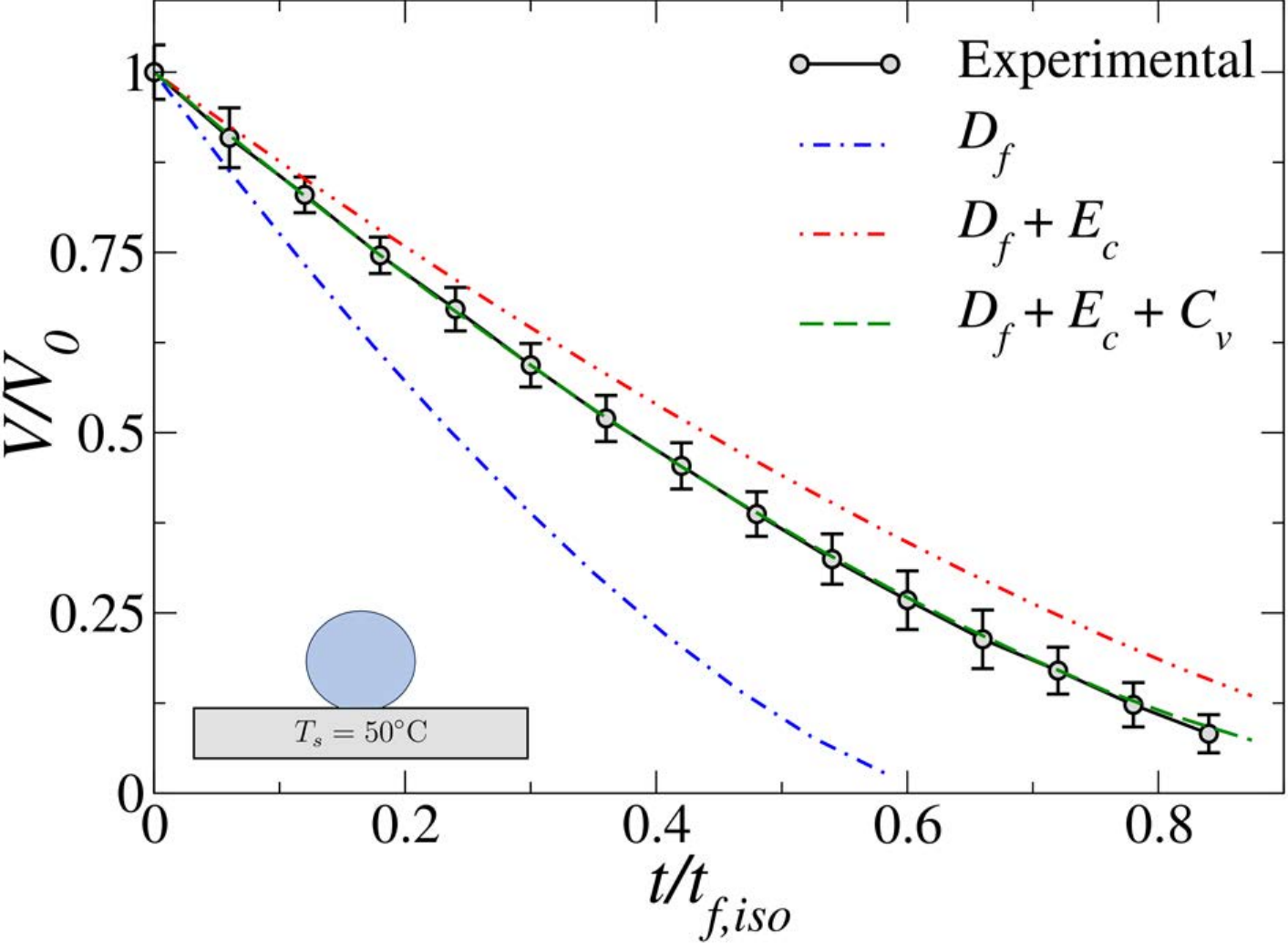} \hspace{2mm} \includegraphics[width=0.46\textwidth]{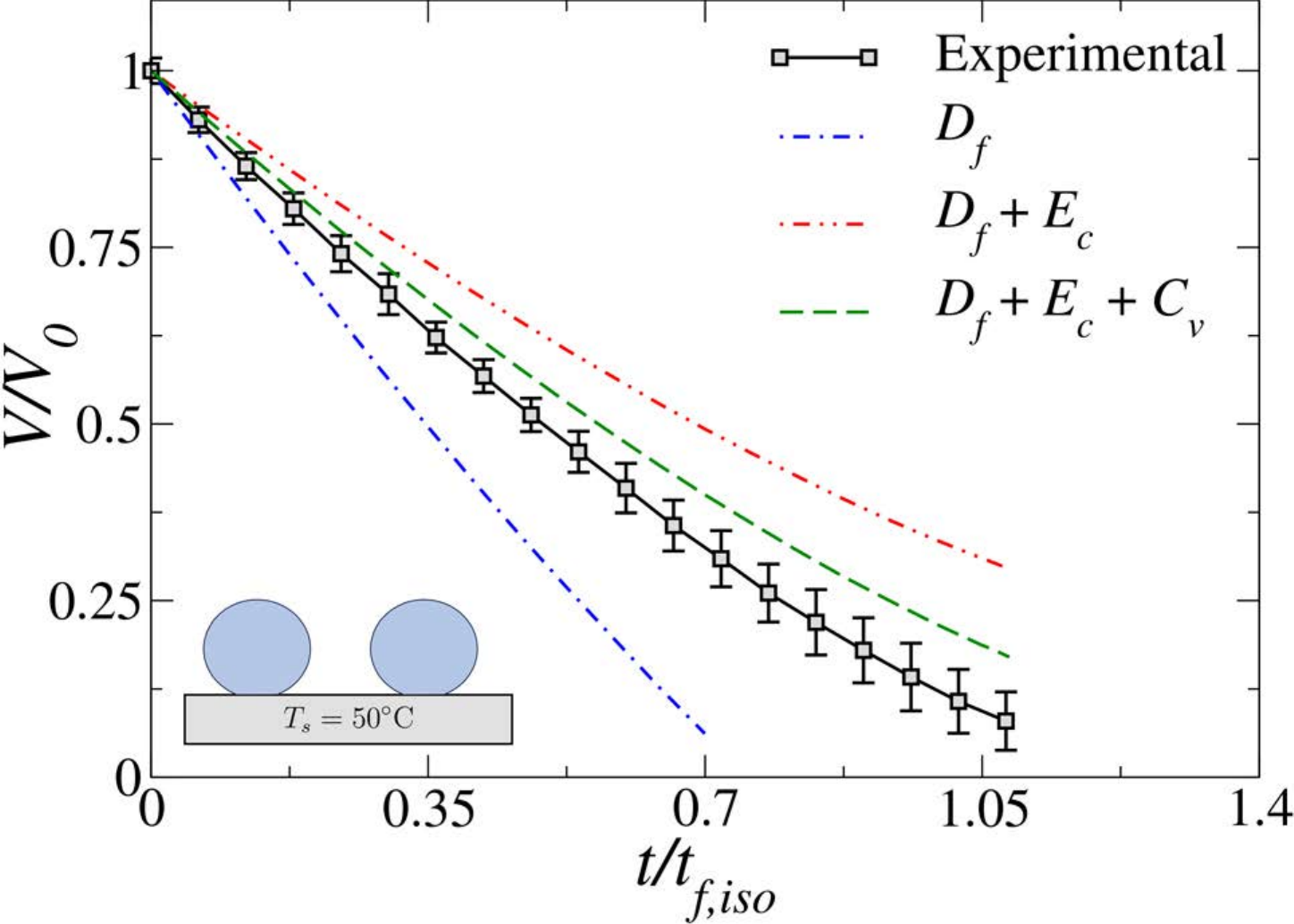}
\caption{Comparison of experimentally measured and theoretically predicted normalized volume ($V/V_0$) as a function of normalized time ($t/t_{f,\mathrm{iso}}$) for (a,c) isolated and (b,d) two-drop configurations. Panels (a,b) correspond to $T_s = 27^\circ$C, and panels (c,d) to $T_s = 50^\circ$C. Error bars on the experimental data represent the standard deviation from three repeated measurements. Theoretical predictions are provided for three models: the purely diffusion-based model ($D_f$), the diffusion with evaporative cooling model ($D_f + E_c$), and the comprehensive model incorporating diffusion, evaporative cooling, and convection ($D_f + E_c + C_v$). In the diffusion-only model ($D_f$), the drop temperature is assumed to be uniform and equal to the substrate temperature $(K(E,\theta) = 1)$.}
\label{fig:fig10}
\end{figure}

To gain deeper insight into the physics of evaporation dynamics for droplets in isolated and two-drop configurations and to predict the temporal variation of the normalized volume during evaporation, we analyze theoreticaly the evaporation process for sessile droplets on superhydrophobic substrates at both room and elevated temperatures. The theoretical model employed in this study is similar to that presented in Refs. \cite{dash2013droplet, dash2014droplet, masoud2021evaporation}. It is well known that the evaporation dynamics of an isolated drop with negligible contact angle hysteresis ($<1^\circ$) exhibits a constant contact angle (CCA) mode, and its evaporation at room temperature can be accurately predicted using the vapor diffusion model. As discussed in the previous section, in our experiments, the isolated droplet primarily followed a constant contact angle (CCA) mode at $T_s = 27^\circ$C. At $T_s = 50^\circ$C, the droplet also exhibited a CCA mode for at least half of its lifetime. For applying the diffusion-based model, we assume that the vapor concentration at the liquid-gas interface, $C_{sat}(T_s)$, corresponds to saturation. The vapor concentration far from the droplet is taken as $RH \times C_{\infty}(T_{\infty})$, where $T_{\infty}$ is the ambient temperature. The rate of evaporation of the isolated droplet based on purely diffusion-based model can be expressed as \cite{dash2014droplet}: 
\begin{equation}\label{diff}
\frac{dm_{iso}}{dt} = - \pi R_{c} D_{v} \left[ C_{sat}(T_s) - RH \times C_{\infty}(T_{\infty}) \right] f(\theta),
\end{equation}
where
\begin{equation}\label{f_theta}
f(\theta) = \frac{\sin\theta}{1 + \cos\theta} + 4 \int_{0}^{\infty} \frac{1 + \cosh(2\theta\tau)}{\sinh(2\pi\tau)} \tanh[(\pi - \theta)\tau] \, d\tau.
\end{equation}
Here, $m_{iso}$ is the mass of the isolated drop, $C_{\infty}$ is the saturation vapor concentration at the ambient temperature ($T_\infty$), $RH$ is the relative humidity, $R_{c}$ is the contact radius, $\theta = \cos^{-1} \left( \frac{R_c^2 - h^2}{R_{c}^2 + h^2} \right)$ is the contact angle calculated from the height ($h$) and the contact diameter ($d_c$) of the drop, and $D_v$ is the diffusion coefficient of water vapor in air at the average temperature between the substrate and the ambient $\left(\frac{T_s + T_{\infty}}{2}\right)$ \cite{gurrala2019evaporation}. For a droplet undergoing evaporation in the CCA mode, using the spherical cap assumption, eqs. (\ref{diff}) and (\ref{f_theta}) can be utilized to obtain the instantaneous volume of the drop ($V$) \cite{dash2014droplet}.
\begin{equation}\label{V_t}
V^{2/3} = V_0^{2/3} - \frac{2\pi D_{v} \left[c_{sat}(T_{s}) - RH \times C_{\infty}(T_{\infty})\right]}{3 \rho_{l}} 
\left(\frac{3}{\pi}\right)^{1/3} \times \left[g(\theta)\right]^{1/3} f(\theta)t,\\
\end{equation}
where
\begin{equation}\label{g_theta}
g(\theta) = \frac{\sin^3\theta}{(1 - \cos\theta)^2 (2 + \cos\theta)}.
\end{equation}
In eq. (\ref{V_t}), $\rho_l$ denotes the density of water. Figure \ref{fig:fig10}(a,b) and \ref{fig:fig10}(c,d) depict the comparison of the experimental observations with the theoretical predictions for the isolated and two-drop configurations at different temperatures, respectively. Figures \ref{fig:fig10}(a,b) and (c,d) correspond to $T_s = 27^\circ$C and $50^\circ$C, respectively. As shown in Figure \ref{fig:fig10}(a), the vapor diffusion model, represented by ($D_f$), adequately predicts the experimental variation of the normalized volume ($V/V_0$) with normalized time ($t/t_{iso}$) for an isolated drop at room temperature ($T_s = 27^\circ$C). It is also noteworthy that the contact angle employed in the calculation ($\theta = 150^\circ$) is close to the average value of the contact angle during evaporation. Changing this $\theta$ value in the range $155^\circ \pm 5^\circ$ does not significantly alter the temporal evolution of the volume predicted by the theory. For the two-drop system at room temperature, we adopt the diffusion-based model for multiple droplets developed by \citet{masoud2021evaporation}. The applicability of this model was tested for three different contact angles ($60^\circ$, $90^\circ$, and $120^\circ$) by \citet{iqtidar2023drying}, and it was found that the model provides very good predictions for $L_c/R_c > 3$; $R_c$ being the wetting radius. In our experiments, the ratio ($L_c/R_c \approx 5.46$) is significantly greater than 3. However, since the surface is superhydrophobic, the minimum distance between the surfaces of the two drops is as small as $0.23$ mm, i.e., $L_e/R_c=0.52$. Assuming the drops in the two-drop system are of equal size and that evaporation is equal on both drops, the evaporation rate of the isolated drop and the drop in the two-drop system can be related by the following expression \cite{masoud2021evaporation, iqtidar2023drying, chen2022predicting}:
\begin{equation}\label{J_exp}
J = \frac{J_{iso}}{1+\phi},
\end{equation}
where, $J_{iso}$ is the evaporation rate of the isolated droplet, $J$ is the evaporation rate of the individual drop in the two-drop system, and $\phi$ is the dimensionless concentration field of the drop in isolation. From the analytical solution obtained for the evaporation of an isolated spherical cap droplet \cite{popov2005evaporative,kek1992dropwise}, the distribution of the dimensionless vapor concentration can be approximated as:
\begin{equation}\label{Pi_exp}
\phi = 4A \frac{R_c}{\tilde{r}} + (A - 4B) \frac{R_{c}^3 \left( \tilde{r}^2 - 3h^2 \right)}{\tilde{r}^5} + O \left[ \left( \frac{R_{c}}{\tilde{r}} \right)^5 \right],
\end{equation}
where
\begin{equation}\label{Con_A}
A = \frac{1}{8\pi R_c} \int_S \mathbf{n} \cdot \nabla \phi \, dS = \int_0^\infty \left\{ \frac{1 + \cosh\left[(2\pi - \theta) \tau\right]}{\cosh(\theta \tau)} \right\}^{-1} d\tau,
\end{equation}
\begin{equation}\label{Con_B}
B = \int_0^\infty \left\{ \frac{1 + \cosh\left[(2\pi - \theta) \tau\right]}{\cosh(\theta \tau)} \right\}^{-1} \tau^2 \, d\tau.
\end{equation}
The evaporation rate of the two-droplet configuration is given by:
\begin{equation}\label{diff_mod_two}
\frac{dm_{two}}{dt} = \frac{- \pi R_{c} D_{v} \left[ C_{sat}(T_s) - RH \times C_{\infty}(T_{\infty}) \right] f(\theta)}{1+\phi}.
\end{equation}
where $\tilde{r} = \sqrt{(h^2 + L_{c}^2)}$, the contact angle can be calculated as $\theta = \cos^{-1} \left( \frac{R_c^2 - h^2}{R_{c}^2 + h^2} \right)$, and $S$ is the free surface. Here, $R_c$ is the contact radius, $D_v$ is the diffusion coefficient. We calculate the evaporation rate of a drop in the two-drop system using the analytical expression for $\phi$. The evaporation rate of an isolated drop can be determined using Eqs.~(\ref{diff}) and (\ref{f_theta}). Then, by employing the relation $J = \rho_l \frac{dV}{dt}$, we integrate $J$ with respect to time to obtain the temporal variation of the remaining volume of the drop. The variation of the normalized volume ($V/V_{0}$) with normalized time ($t/t_{f,\text{iso}}$) for the two-drop system, as obtained from experiments and the diffusion-based theoretical model ($D_f$) at $T_s = 27^\circ$C, indicates that the model underestimates the time required for the droplet volume to reduce to half its initial value ($V/V_0 = 0.5$) by approximately 18\% (Figure~\ref{fig:fig10}b). However, for the isolated drop on the substrate maintained at $50^\circ$C, the vapor diffusion model ($D_f$) underpredicts the evaporation time by approximately $37\%$ as the drop reaches half its volume ($V/V_{0} = 0.5$), as seen in Figure \ref{fig:fig10}(c). It is important to note that on superhydrophobic substrates at elevated temperatures, the enhancement in evaporation due to convection in the air and liquid phases is insufficient to compensate for the suppression caused by interfacial cooling resulting from evaporation \cite{dash2014droplet}. To improve the accuracy of the prediction, we incorporate the correction factor proposed by \citet{shen2022numerical}, which accounts for evaporative cooling by coupling the temperature and vapor concentration at the liquid–air interface. This correction factor helps in predicting the actual evaporation rate for the non-isothermal system from the isothermal formulation (eqs. \ref{diff}). The correction factor ($K(E,\theta)$) introduced by \citet{shen2022numerical} is applicable only when the temperature difference between the substrate and the ambient satisfies $T_s - T_{\infty} \geq 7^\circ$C. This factor depends on the contact angle $\theta$ and the evaporative cooling number $E = \frac{D_v b L_v}{k_l}$, where $b = \frac{dC_{sat}}{dT} \approx \frac{C_{sat} - C_{\infty}}{T_{sat} - T_{\infty}}$. In this expression, $D_v$ is the diffusion coefficient of vapor in air, $L_v$ is the latent heat of vaporization, and $k_l$ is the thermal conductivity of water. The saturation concentration $C_{sat}$ is calculated at the average temperature between the substrate and the ambient $\left(\frac{T_s + T_{\infty}}{2}\right)$, while $C_{\infty}$ is calculated at the ambient temperature. In the theoretical model considered by \citet{nguyen2018analytical,shen2022numerical}, they assumed saturated conditions in the environment and that the saturation concentration varies linearly within the gas domain. All the values of $D_v$, $L_v$, and $k_l$ are taken at the average temperature between the substrate and the ambient $\left({(T_s + T_{\infty})/2}\right)$. The correlation for the correction factor ($K(E,\theta)$) is provided in Refs. \cite{erbil2023droplet, shen2022numerical, jenkins2023suppression} (also see, eqs. (S1) and (S2) in supplementary material). For the present system, with $E = 0.19$ and $\theta = 150^\circ$, the value of $K(E,\theta)$ is calculated to be $0.54$. However, this correction factor ($K(E,\theta)$) cannot be applied when the substrate and ambient are at room temperature, as the correlation for $K(E,\theta)$ (provided in equations (S1) and (S2) of the supplementary material) is valid only for cases where the temperature difference satisfies $(T_s - T_{\infty}) \geq 7^\circ$C. Therefore, we do not employ this correlation for the isolated and two-drop systems at room temperature.

Based on the diffusion plus evaporative cooling model ($D_f + E_c$), the evaporation rate of the drop is given by:
\begin{equation}\label{diff_mod}
\frac{dm_{iso}}{dt} = - \pi K(E,\theta)R_{c} D_{v} \left[ C_{sat}(T_s) - RH \times C_{\infty}(T_{\infty}) \right] f(\theta),
\end{equation}
The instantaneous volume of the drop is given by:
\begin{equation}\label{V_t_mod}
V^{2/3} = V_0^{2/3} - \frac{2\pi D_{v} K(E,\theta)\left[c_{sat}(T_{s}) - RH \times C_{\infty}(T_{\infty})\right]}{3 \rho_{l}} 
\left(\frac{3}{\pi}\right)^{1/3} \times \left[g(\theta)\right]^{1/3} f(\theta)t.
\end{equation}

In Figure~\ref{fig:fig10}(c), it is observed that the normalized droplet volume ($V/V_0$), as predicted by the modified theoretical model (Eq.~\ref{V_t_mod}) that accounts for evaporative cooling ($D_f + E_c$), overestimates the experimentally measured evaporation time by approximately 16\% when the droplet volume reduces to half of its initial value ($V/V_0 = 0.5$). It is to be noted that for the isothermal model (Eq. \ref{diff}) when both the substrate and ambient environment are at room temperature, $K(E,\theta) = 1$ \cite{shen2022numerical}. As buoyancy-driven convection plays a significant role in enhancing evaporation at elevated temperatures \cite{dash2014droplet}, we adopt the empirical correlation proposed by \citet{kelly2011evaporation}. This model incorporates the effects of buoyant convection through the Grashof number ($Gr$), defined as
\begin{equation}\label{corr} 
Gr = \frac{P_v M g R_{c}^{3}}{(P_a - P_v) M_a \nu_a^{2}}, 
\end{equation}
and relates it to the enhancement factor of evaporation,
\begin{equation}\label{corr_f} 
E_r = 0.31Gr^{0.216}, 
\end{equation}
where $E_r$ is the ratio of the evaporation rate due to convection to that due to diffusion. Here, $P_v$ is the saturation vapor pressure, $P_a$ is the atmospheric pressure, $M$ and $M_a$ are the molecular weights of water vapor and air, respectively, $\nu_a$ is the kinematic viscosity of air, and $R_c$ is the contact radius of the drop. This empirical correlation effectively captures the influence of vapor concentration, droplet size, and substrate temperature on the evaporation dynamics. Using this relationship, the combined diffusion+evaporative cooling+convection model ($D_f + E_c + C_v$) can be expressed as:
\begin{align}\label{E_conv}
\frac{dm}{dt} &= - \pi R_{c} K(E,\theta)D_{v} \left[ C_{sat}(T_s) - RH \times C_{\infty}(T_{\infty}) \right] f(\theta)  \times \left[1 + 0.31Gr^{0.216}\right].
\end{align}
Here, $R_c$ denotes the contact radius, and $D_v$ is the diffusion coefficient. The evaporation rate of an isolated drop at elevated substrate temperature can be determined using Eqs.~(\ref{E_conv}) and (\ref{f_theta}). It can be seen in Figure~\ref{fig:fig10}(c) that the modified model (Eq. \ref{E_conv}), which accounts for natural convection and evaporative cooling ($D_f + E_c + C_v$), provides theoretical predictions that agree well with the experimental results for an isolated droplet at elevated temperature ($T_s = 50^\circ$C). We further use Eq.  (\ref{E_conv}) to calculate the evaporation rate of a drop in the two-drop system at elevated temperatures using the analytical expression for dimensionless concentration field $\phi$ (Eq. \ref{Pi_exp}). For the two-drop system at elevated temperature the combined diffusion+evaporative cooling+convection model ($D_f + E_c + C_v$) can be expressed as:
\begin{align}\label{E_conv_two}
\frac{dm}{dt} &= \frac{- \pi R_{c} K(E,\theta)D_{v} \left[ C_{sat}(T_s) - RH \times C_{\infty}(T_{\infty}) \right] f(\theta)}{1+\phi}  \times \left[1 + 0.31Gr^{0.216}\right].
\end{align}
Then, by employing the relation $J = \rho_l \frac{dV}{dt}$, we integrate $J$ with respect to time to obtain the temporal variation of the remaining volume of the drop. For the two-drop system at elevated temperature in Figure \ref{fig:fig10}(d), for $T_s = 50^\circ$C, we observe that the purely diffusion-based model $(D_f)$ underpredicts the time for the droplet volume to reduce to half of its initial value by approximately $31\%$, whereas the diffusion plus evaporative cooling model ($D_f + E_c$) overpredicts it by $39\%$. In contrast, the combined diffusion+evaporative cooling+convection model ($D_f + E_c + C_v$), described by Eq.~(\ref{E_conv_two}), shows very good agreement with the experimental data, overestimating the time for the droplet volume to reach half its initial value ($V/V_0 = 0.5$) by only 14\%. This discrepancy between the experimental results and theoretical predictions arises because the diffusion-based model does not account for evaporation-induced cooling in both the isolated drop and two-drop systems, nor does it capture the close spacing between the surfaces of the two drops in the two-drop system ($L_e \leq 0.37$ mm). The reduced spacing between the drops leads to significant vapor accumulation, which is not fully captured by the model due to the underlying spherical cap assumption, resulting in the observed deviations. The correction factor ($K(E,\theta)$) derived by \citet{shen2022numerical} applies to an isolated drop under saturation conditions, which differs substantially from the conditions of the two-drop scenarios considered in the present study. Furthermore, the dimensionless concentration field in Eq. (\ref{Pi_exp}) used in the model only accounts for changes in vapour diffusion due to the presence of nearby drops. In contrast, the factor accounting for convection in Eq. (\ref{corr_f}) corresponds to a single-drop system with low contact angles, in contrast to the high contact angle observed on superhydrophobic substrates. These results demonstrate that the theoretical model incorporating all three effects ($D_f + E_c + C_v$) effectively captures the experimental behavior for both isolated and single-drop systems at elevated temperatures. In contrast, the pure diffusion-based model ($D_f$) shows the closest agreement with experimental observations at room temperature.

\section{Conclusion}\label{sec:conc}
The present study provides a comprehensive analysis of the evaporation dynamics of two droplets placed in close proximity on micro-nano textured superhydrophobic substrates at both room temperature ($T_s = 27^\circ$C) and elevated temperatures ($T_s = 50^\circ$C), utilizing the shadowgraphy imaging technique. The side-view profiles of the droplets are extracted through post-processing of images captured by a CMOS camera, using a custom program developed in the \textsc{Matlab}$^{\circledR}$ framework. The morphology of the droplets shows that the two-drop system evaporates more slowly than isolated droplets due to the vapour shielding effect, which increases the vapour concentration between the droplets, leading to longer lifetimes and asymmetric evaporation. At room temperature, the lifetime of the two-drop system is approximately $1.6$ times that of the isolated droplet, while at the elevated temperature of $50^\circ$C, it is $1.2$ times longer. At room temperature, isolated droplets primarily follow a constant contact angle (CCA) mode, with occasional stick-slip events, while the two-drop system predominantly exhibits CCA mode, with intermittent phases of mixed-mode evaporation. At elevated temperatures, the isolated droplet transitions from a constant contact angle to a mixed evaporation mode, whereas the two-drop system follows a mixed-mode evaporation throughout its lifetime. The temporal variations of normalized wetting diameter ($d/d_{c0}$), contact angle ($\theta$), and normalized volume ($V/V_0$) of both the isolated droplet and the two-drop system confirm these evaporation behaviors. A diffusion-based theoretical model has been developed to enhance the understanding of evaporation behavior in isolated and two-droplet configurations under both room and elevated temperature conditions. Although the diffusion-based theoretical model aligns well with experimental observations for both isolated and two-drop configurations at room temperature, it fails to accurately capture the evaporation behavior at elevated temperatures. For isolated droplets at higher temperatures, incorporating evaporative cooling and natural convection into the diffusion model significantly improves agreement with experiments. This more comprehensive model, which accounts for diffusion, evaporative cooling, and convection, also provides more accurate predictions for the two-drop system under elevated thermal conditions. To the best of our knowledge, this study is the first to explore the evaporation dynamics of two droplets on superhydrophobic substrates at elevated temperatures, providing novel insights into the intricate interactions of multi-droplet evaporation in confined systems.\\

\noindent{\bf Credit authorship contribution statement} 

S.K. performed the experiments. All of the authors contributed to the analysis of the results and to the preparation of the manuscript. 

\noindent{\bf Declaration of Competing Interest} 

The authors declare that there is no conflict of interest.

\noindent{\bf Supplementary material}
\begin{itemize}

\item Figure S1: The variation of the normalized wetting diameter ($d_c/d_{c0}$) with $t/t_{f,iso}$ obtained in different repetitions for (a) the isolated droplet configuration and (b) the two-drop configuration at $T_s = 27^\circ$C. The corresponding variation of the contact angle ($\theta$) with $t/t_{f,iso}$ is presented for (c) the isolated droplet configuration and (d) the two-drop configuration. The chamber is maintained at a relative humidity of $RH = 0.16$, with an evaporation time of $t_{f,iso} = 1269$ seconds.

\item Figure S2: (a) Demonstration of the left end ($x_{cl}$) and the right end ($x_{cr}$) of the contact line.
(b) Evolution of the left end ($x_{cl}$) and right end ($x_{cr}$) of the contact line as a function of $t/t_{f,iso}$ for an isolated droplet at $T_s = 27^\circ$C. Here, $x_{cl}$ and $x_{cr}$ are measured relative to the droplet centroid position at $t = 0$.

\item Figure S3: The variation of the normalized wetting diameter ($d_c/d_{c0}$) with $t/t_{f,iso}$ obtained in different repetitions for (a) the isolated droplet configuration and (b) the two-drop configuration at at $T_s = 50^\circ$C. The corresponding variation of the contact angle ($\theta$) with $t/t_{f,iso}$ is presented for (c) the isolated droplet configuration and (d) the two-drop configuration. The chamber is maintained at a relative humidity of $RH = 0.16$, with an evaporation time of $t_{f,iso} = 494$ seconds.

\item Figure S4: Temporal variation of the (a,c) normalized wetting diameter ($d_c/d_{c0}$) and (b,d) contact angle ($\theta$) at different substrate temperatures. Panels (a,b) correspond to the isolated droplet configuration, while panels (c,d) correspond to the two-drop configuration. The total evaporation time of the isolated droplet, denoted by $t_{f,\mathrm{iso}}$, is 1280 for $T_s = 27^\circ$C and 498 for $T_s = 50^\circ$C.

\item Calculation of the correction factor $K(E,\theta)$.

\end{itemize}

\noindent{\bf Acknowledgement:} {We thank Prof. Gilbert Walker, Editor-in-Chief of Langmuir, for inviting K.C.S. to submit an article as one of its Editorial Board Members. S.K. extends gratitude to Gopal Chandra Pal, Anand S., and Varun Chaturvedi from IIT Ropar, and Hari Govindha A. from IIT Hyderabad, for insightful discussions on both the experimental and theoretical aspects.}


\providecommand{\latin}[1]{#1}
\makeatletter
\providecommand{\doi}
  {\begingroup\let\do\@makeother\dospecials
  \catcode`\{=1 \catcode`\}=2 \doi@aux}
\providecommand{\doi@aux}[1]{\endgroup\texttt{#1}}
\makeatother
\providecommand*\mcitethebibliography{\thebibliography}
\csname @ifundefined\endcsname{endmcitethebibliography}
  {\let\endmcitethebibliography\endthebibliography}{}

\clearpage

\underline{Supplementary Information} 

\begin{figure}[h]
\centering
\hspace{0.75cm}{\large (a)} \hspace{6.75cm}  {\large (b)} \\
\includegraphics[width=0.45\textwidth]{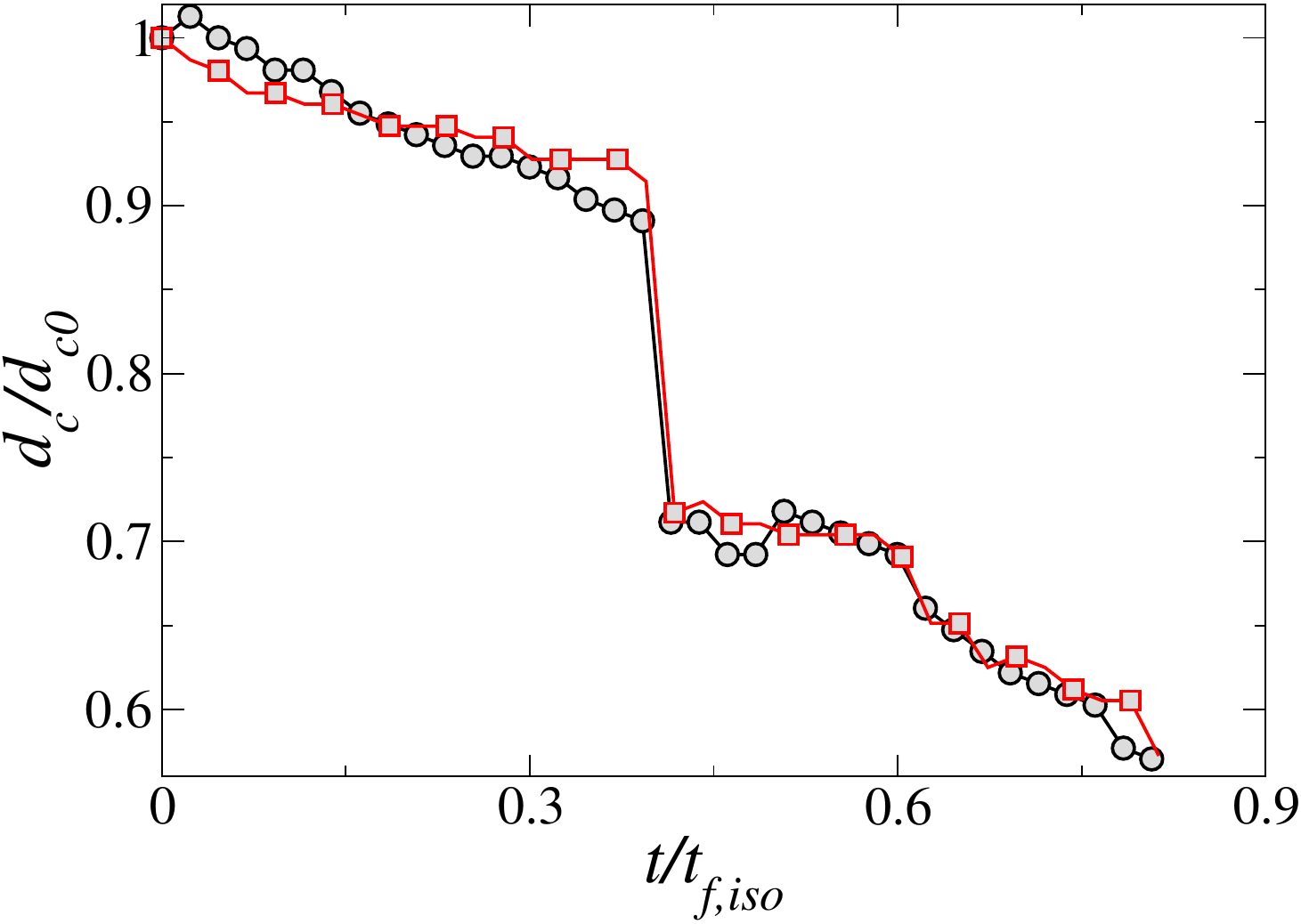} 
\hspace{0mm}
\includegraphics[width=0.45\textwidth]{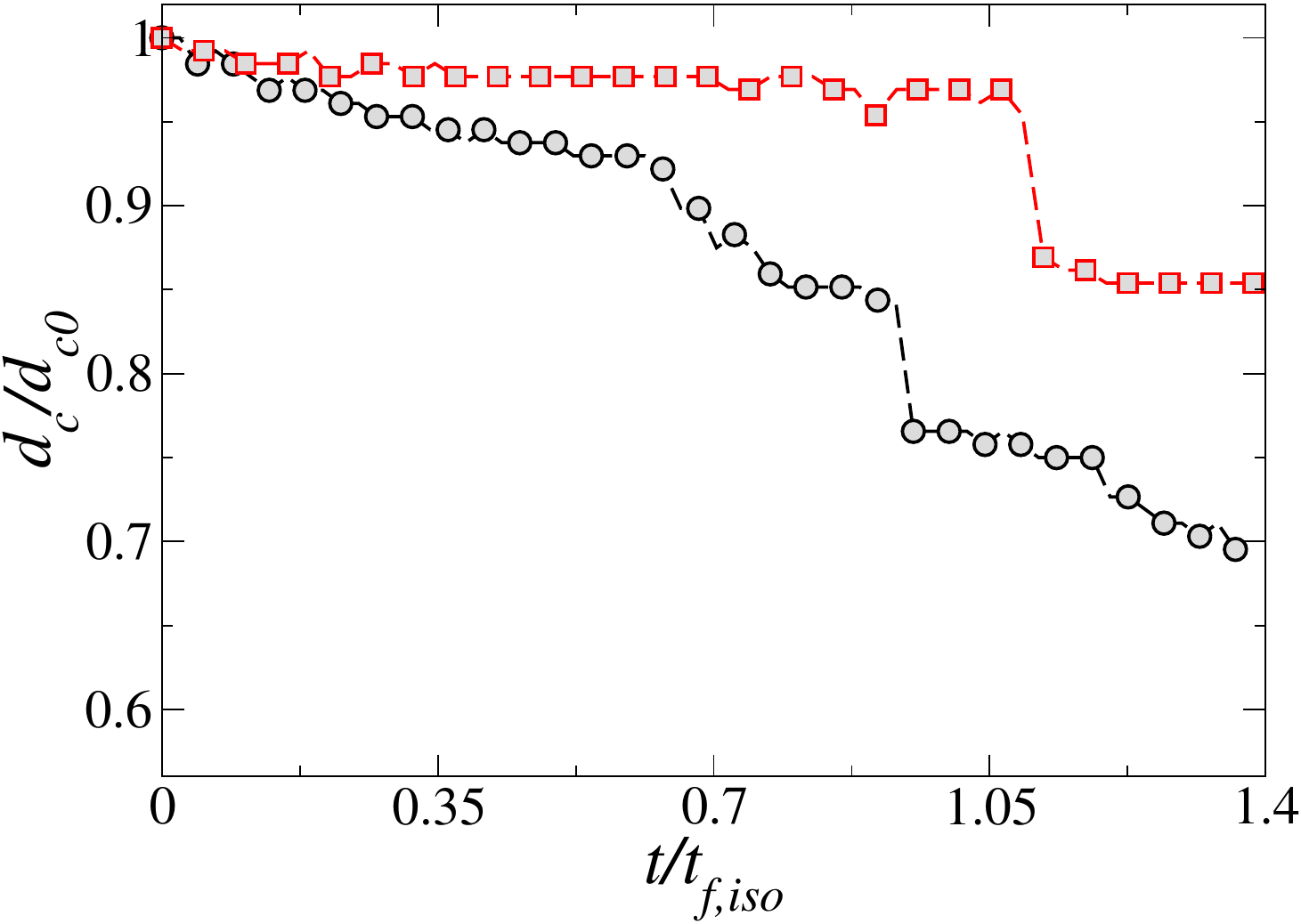}\\
\hspace{0.75cm}{\large (c)}   \hspace{6.75cm}  {\large (d)} \\
\includegraphics[width=0.45\textwidth]{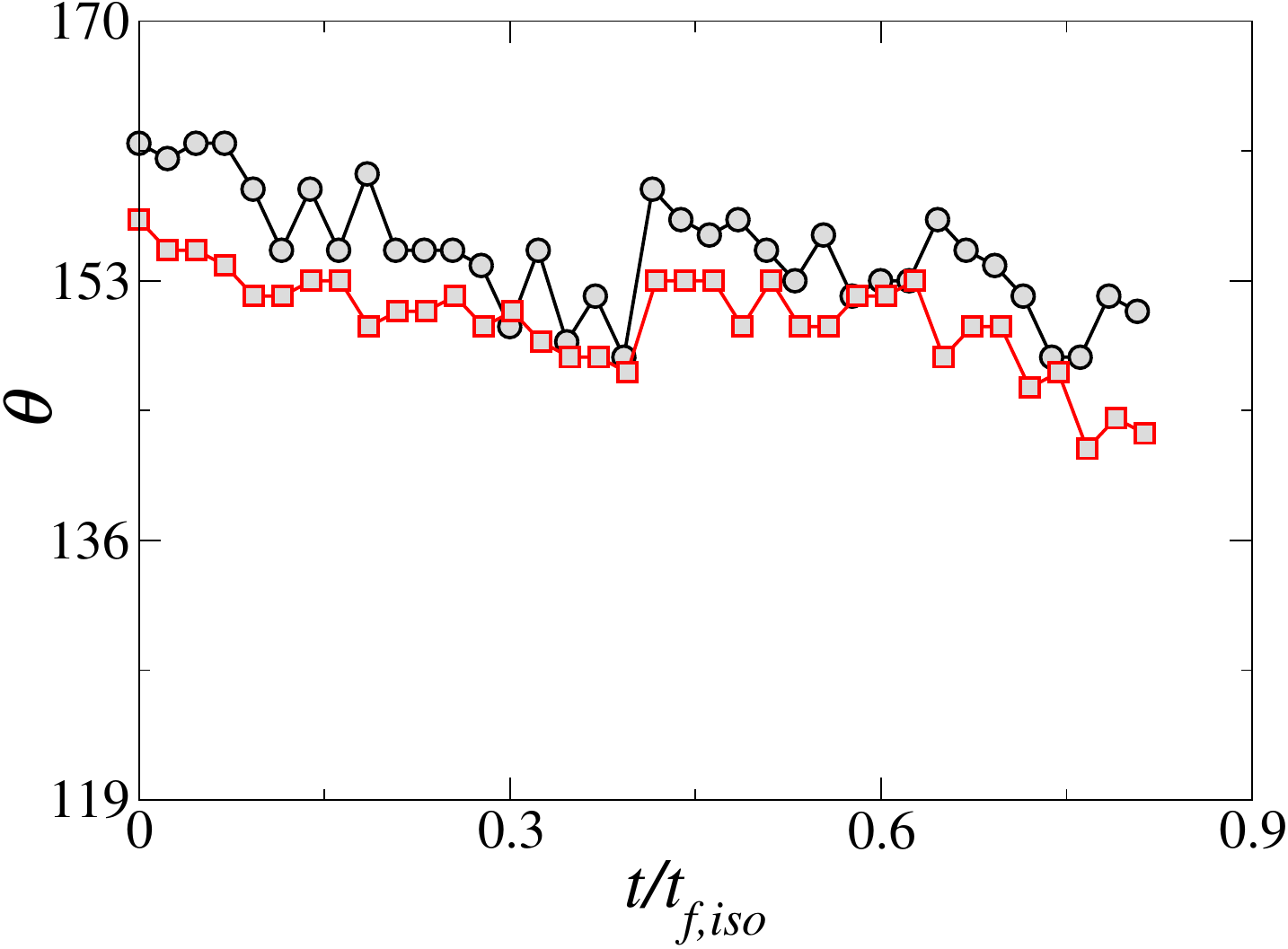}
\hspace{0mm}
\includegraphics[width=0.45\textwidth]{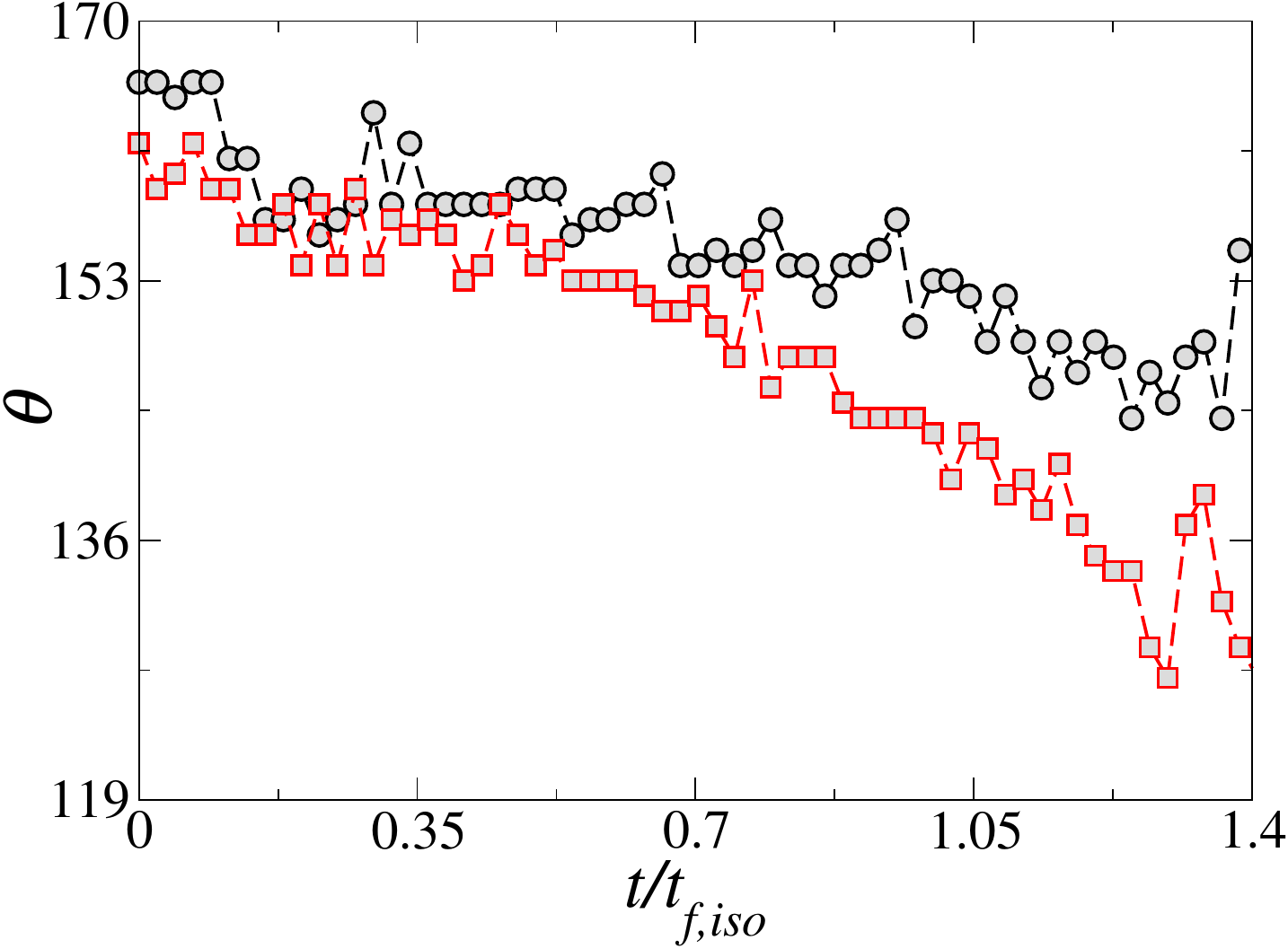}\\
Figure S1: The variation of the normalized wetting diameter ($d_c/d_{c0}$) with $t/t_{f,iso}$ obtained in different repetitions for (a) the isolated droplet configuration and (b) the two-drop configuration at $T_s = 27^\circ$C. The corresponding variation of the contact angle ($\theta$) with $t/t_{f,iso}$ is presented for (c) the isolated droplet configuration and (d) the two-drop configuration. The chamber is maintained at a relative humidity of $RH = 0.16$, with an evaporation time of $t_{f,iso} = 1269$ seconds.
\label{fig:figS1}
\end{figure}

\begin{figure}[h]
\centering
\hspace{0.5cm}  {\large (a)} \hspace{8 cm} {\large (b)} \\
\includegraphics[width=0.45\textwidth]{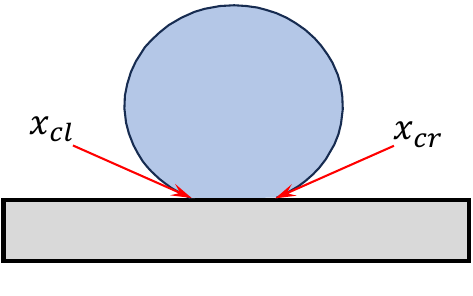} \hspace{2mm}  \includegraphics[width=0.45\textwidth]{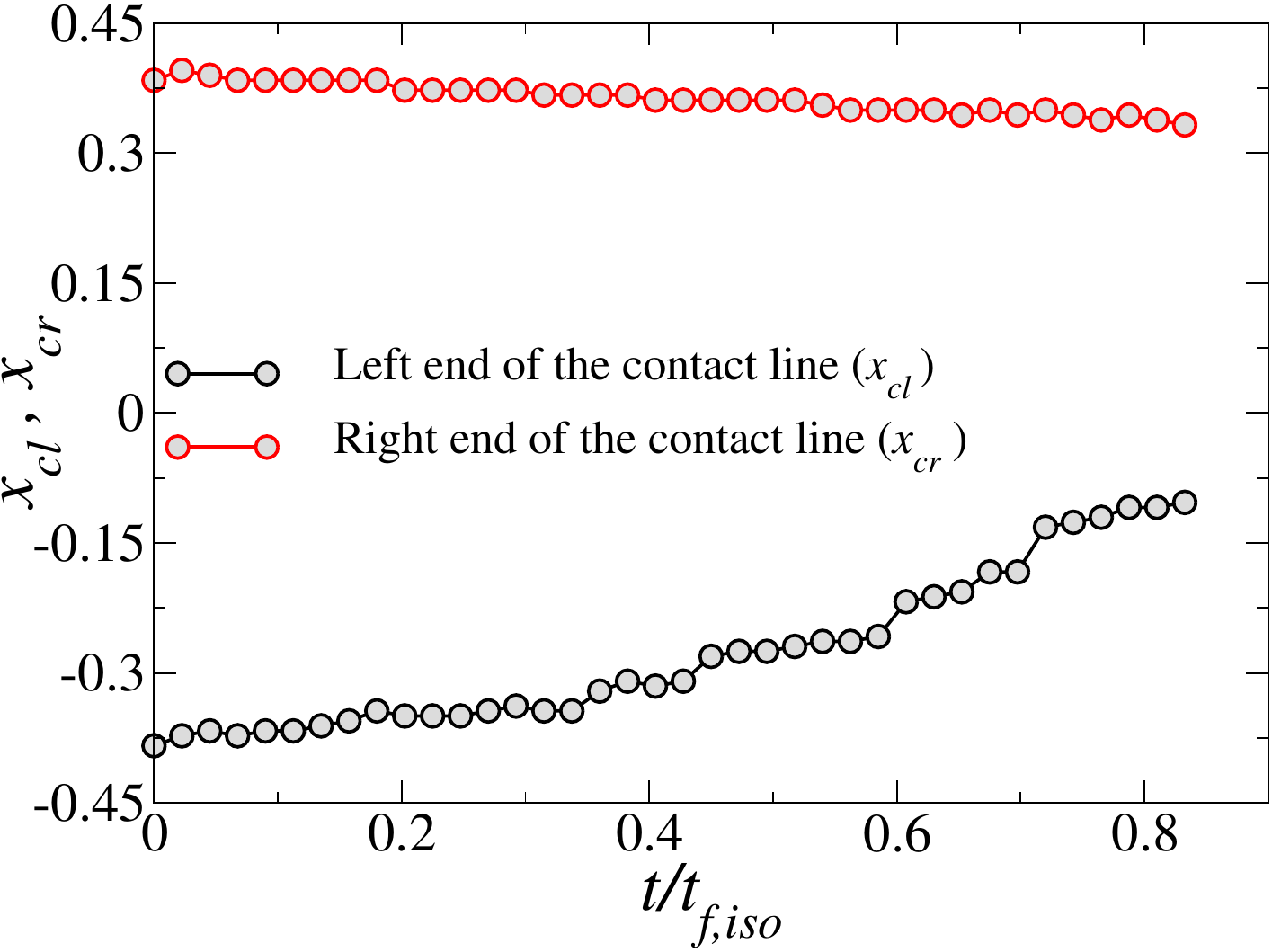}\\
Figure S2: (a) Demonstration of the left end ($x_{cl}$) and the right end ($x_{cr}$) of the contact line.
(b) Evolution of the left end ($x_{cl}$) and right end ($x_{cr}$) of the contact line as a function of $t/t_{f,iso}$ for an isolated droplet at $T_s = 27^\circ$C. Here, $x_{cl}$ and $x_{cr}$ are measured relative to the droplet centroid position at $t = 0$.
\label{fig:xcl_xcr}
\end{figure}

\begin{figure}[h]
\centering
\hspace{0.75cm}{\large (a)} \hspace{6.75cm}  {\large (b)} \\
\includegraphics[width=0.45\textwidth]{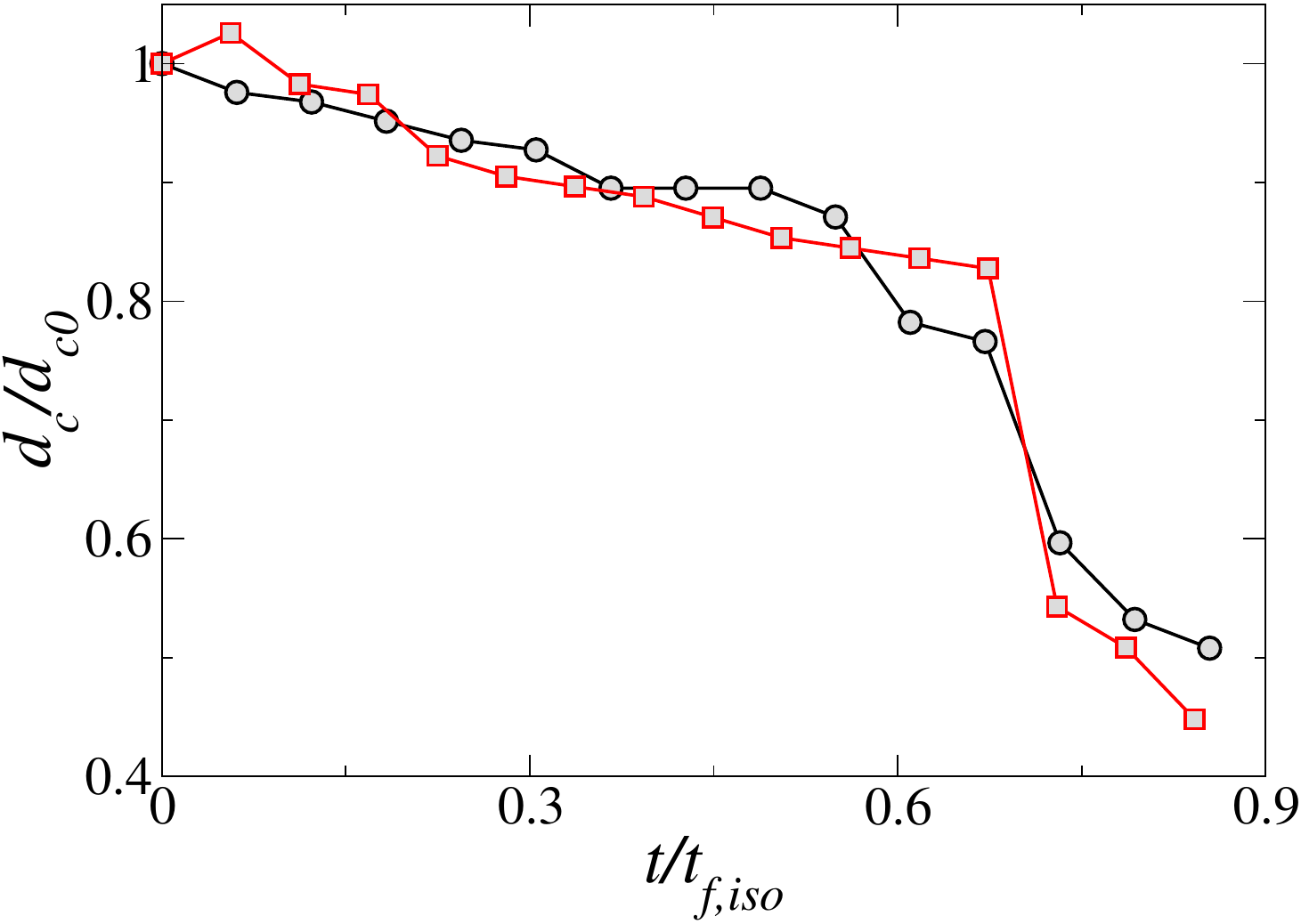} 
\hspace{0mm}
\includegraphics[width=0.45\textwidth]{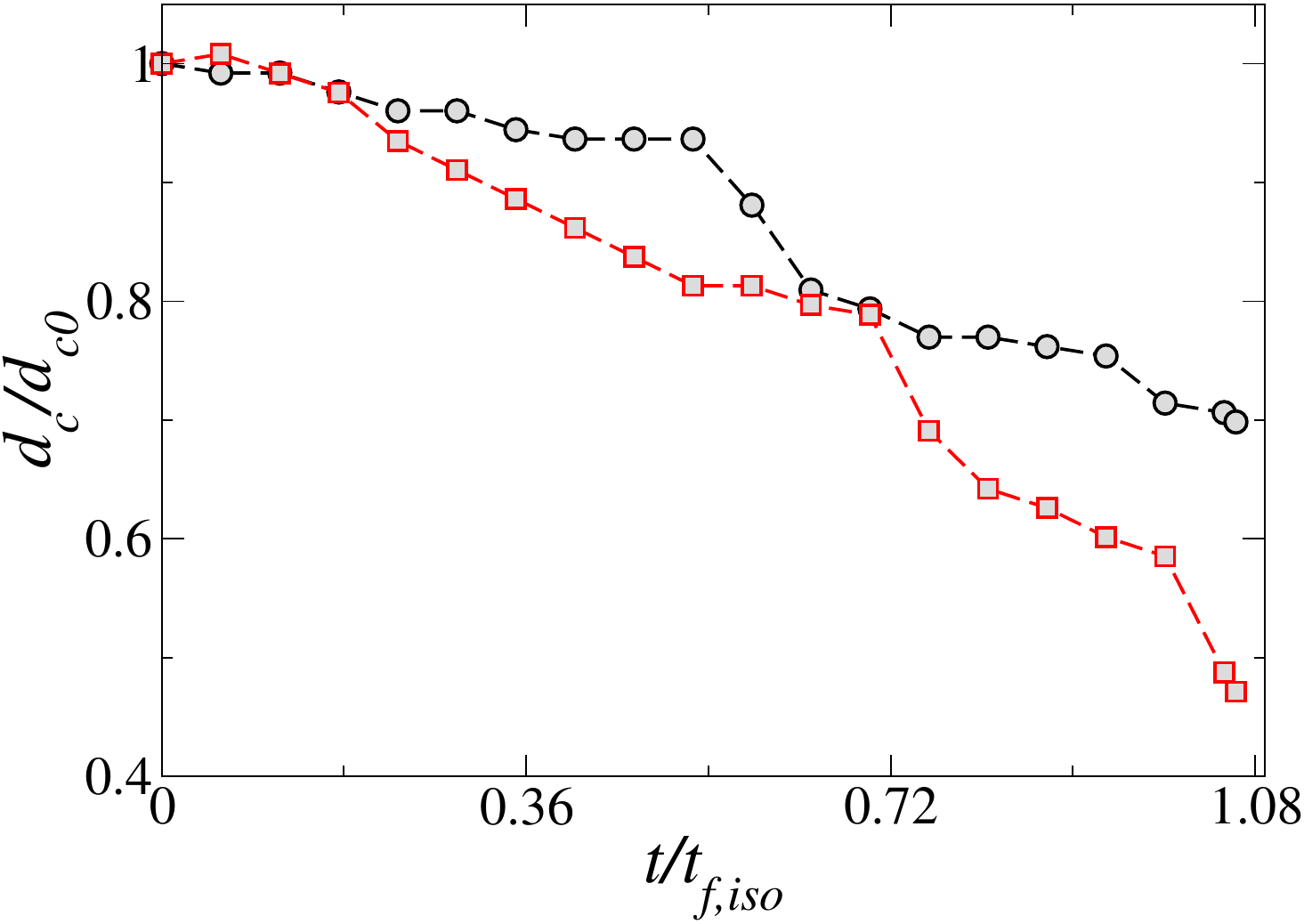}\\
\hspace{0.75cm}{\large (c)}   \hspace{6.75cm}  {\large (d)} \\
\includegraphics[width=0.45\textwidth]{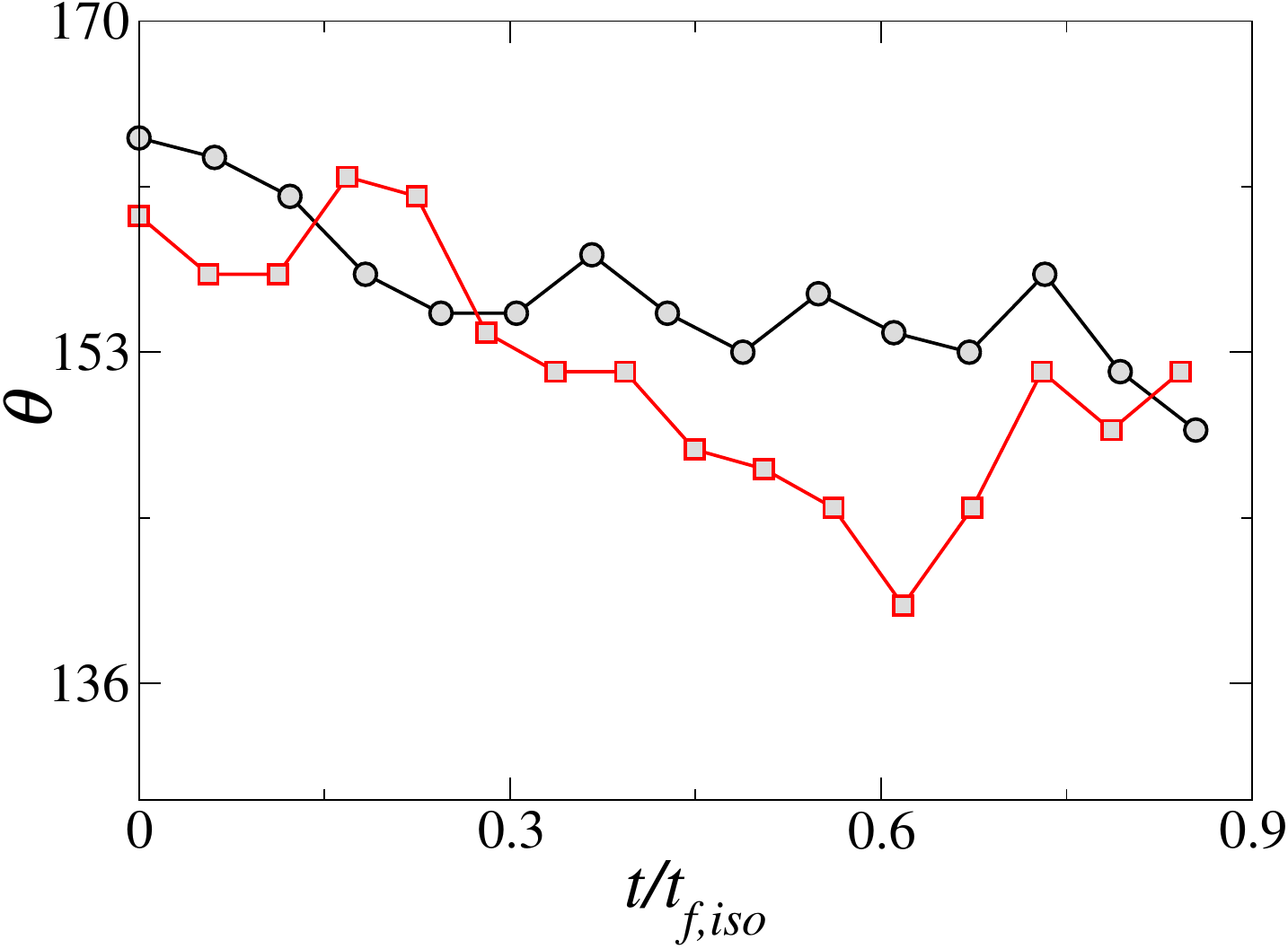}
\hspace{0mm}
\includegraphics[width=0.45\textwidth]{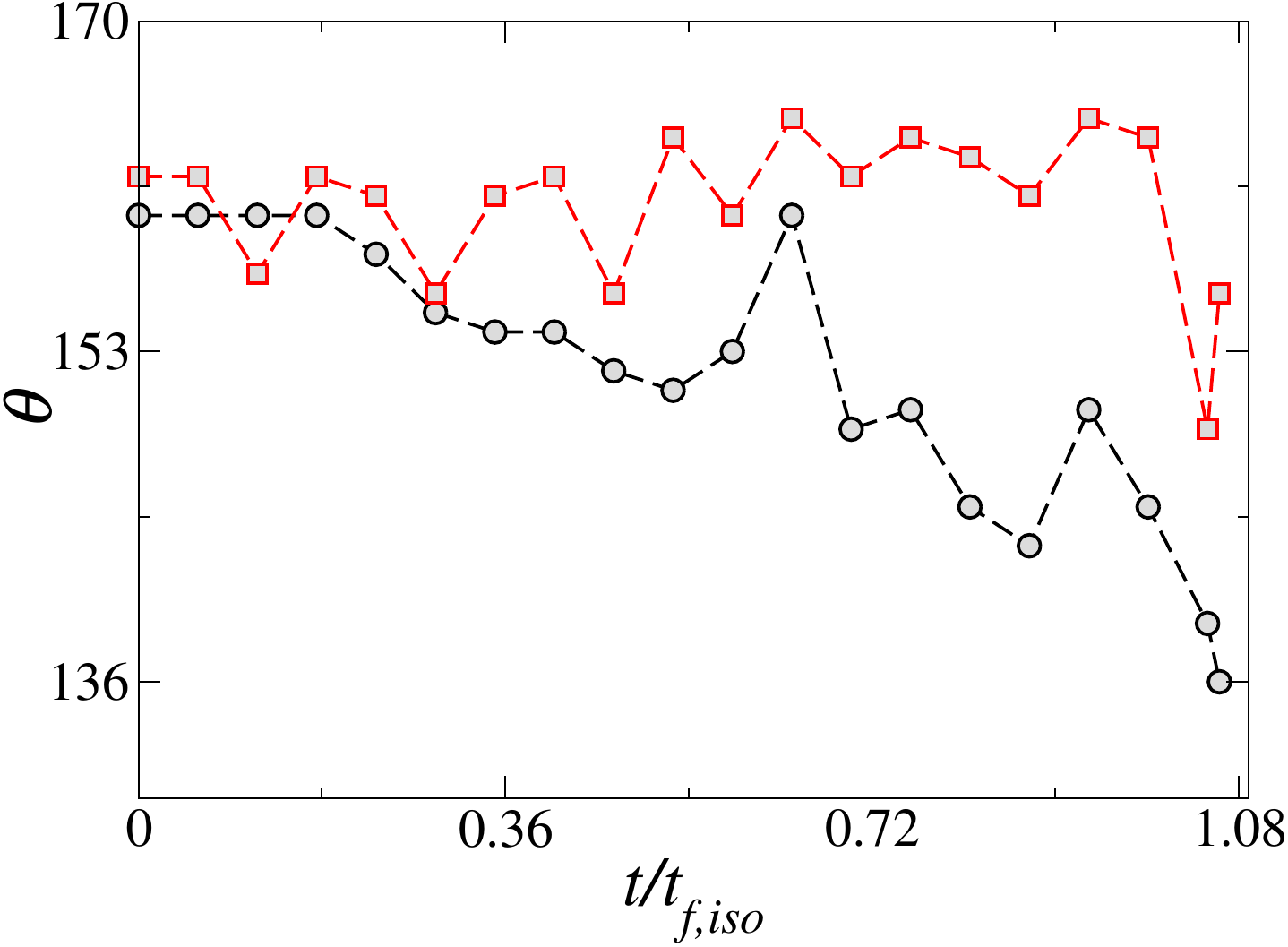}\\
Figure S3: The variation of the normalized wetting diameter ($d_c/d_{c0}$) with $t/t_{f,iso}$ obtained in different repetitions for (a) the isolated droplet configuration and (b) the two-drop configuration at at $T_s = 50^\circ$C. The corresponding variation of the contact angle ($\theta$) with $t/t_{f,iso}$ is presented for (c) the isolated droplet configuration and (d) the two-drop configuration. The chamber is maintained at a relative humidity of $RH = 0.16$, with an evaporation time of $t_{f,iso} = 494$ seconds.
\label{fig:figS2}
\end{figure}

\begin{figure}[h]
\centering
 \hspace{0.5cm}  {\large (a)} \hspace{7.2cm} {\large (b)} \\
 \includegraphics[width=0.45\textwidth]{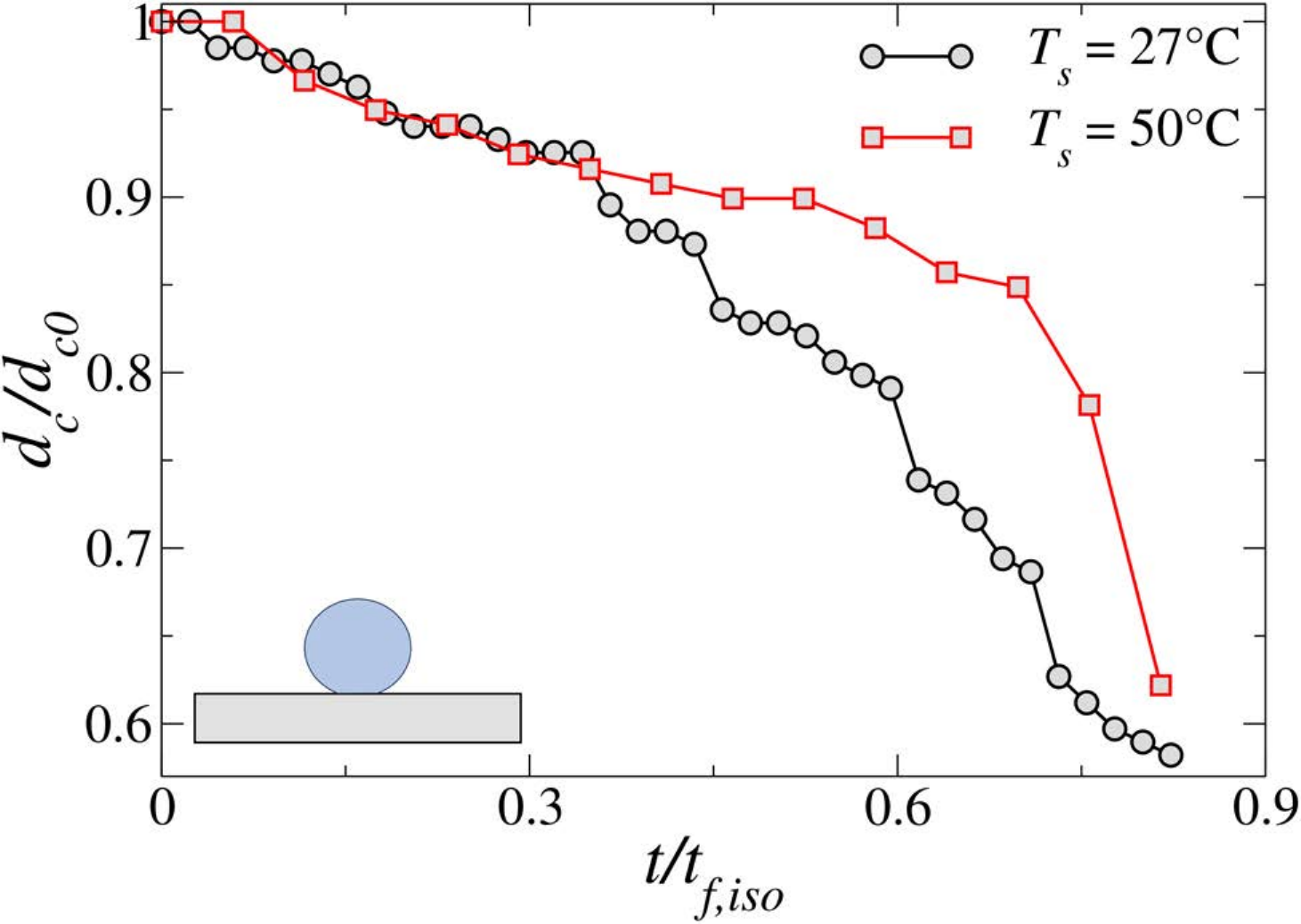} \hspace{2mm} \includegraphics[width=0.45\textwidth]{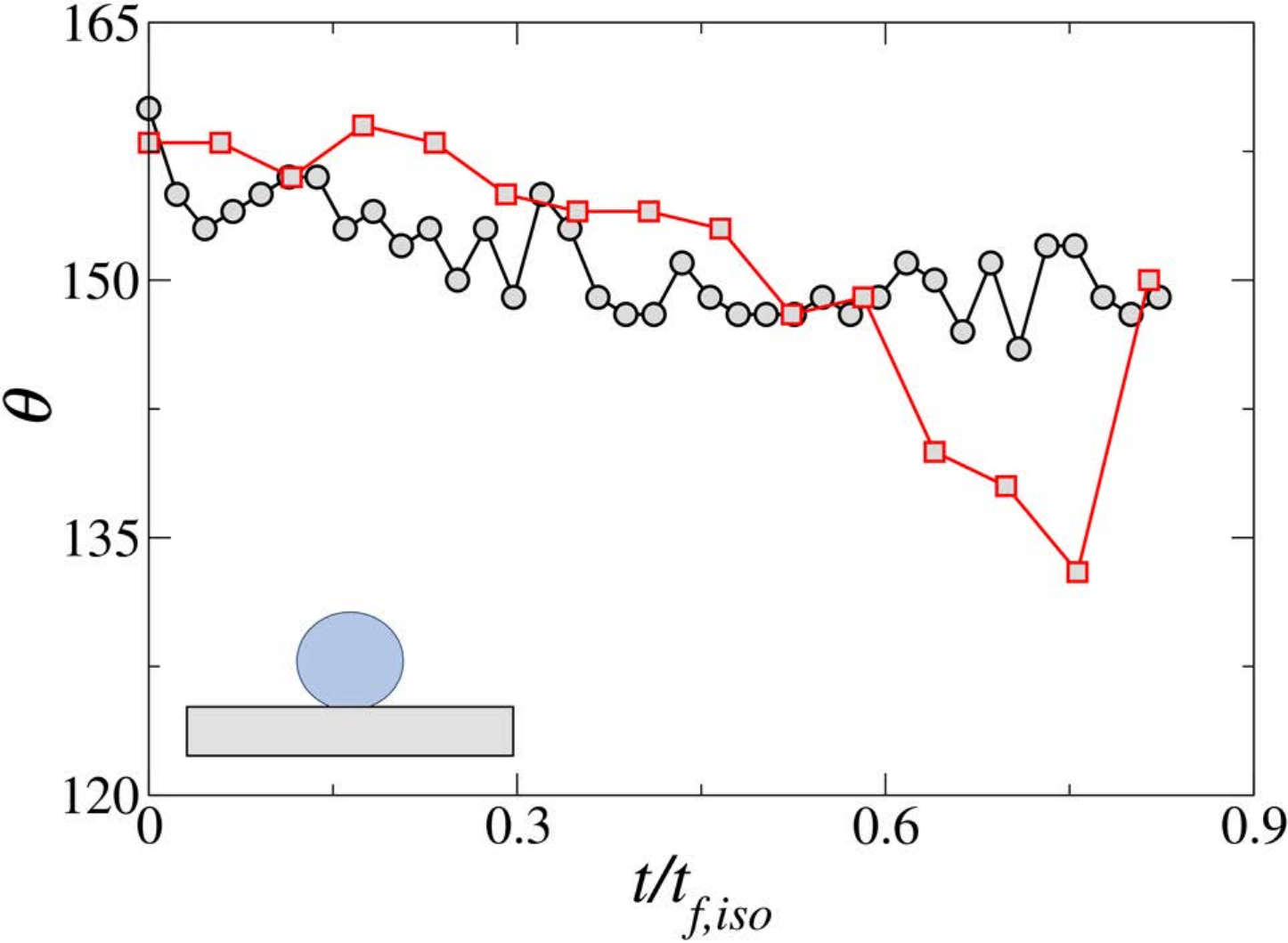}\\
  \hspace{0.5 cm}  {\large (c)} \hspace{7.2cm} {\large (d)} \\
 \includegraphics[width=0.46\textwidth]{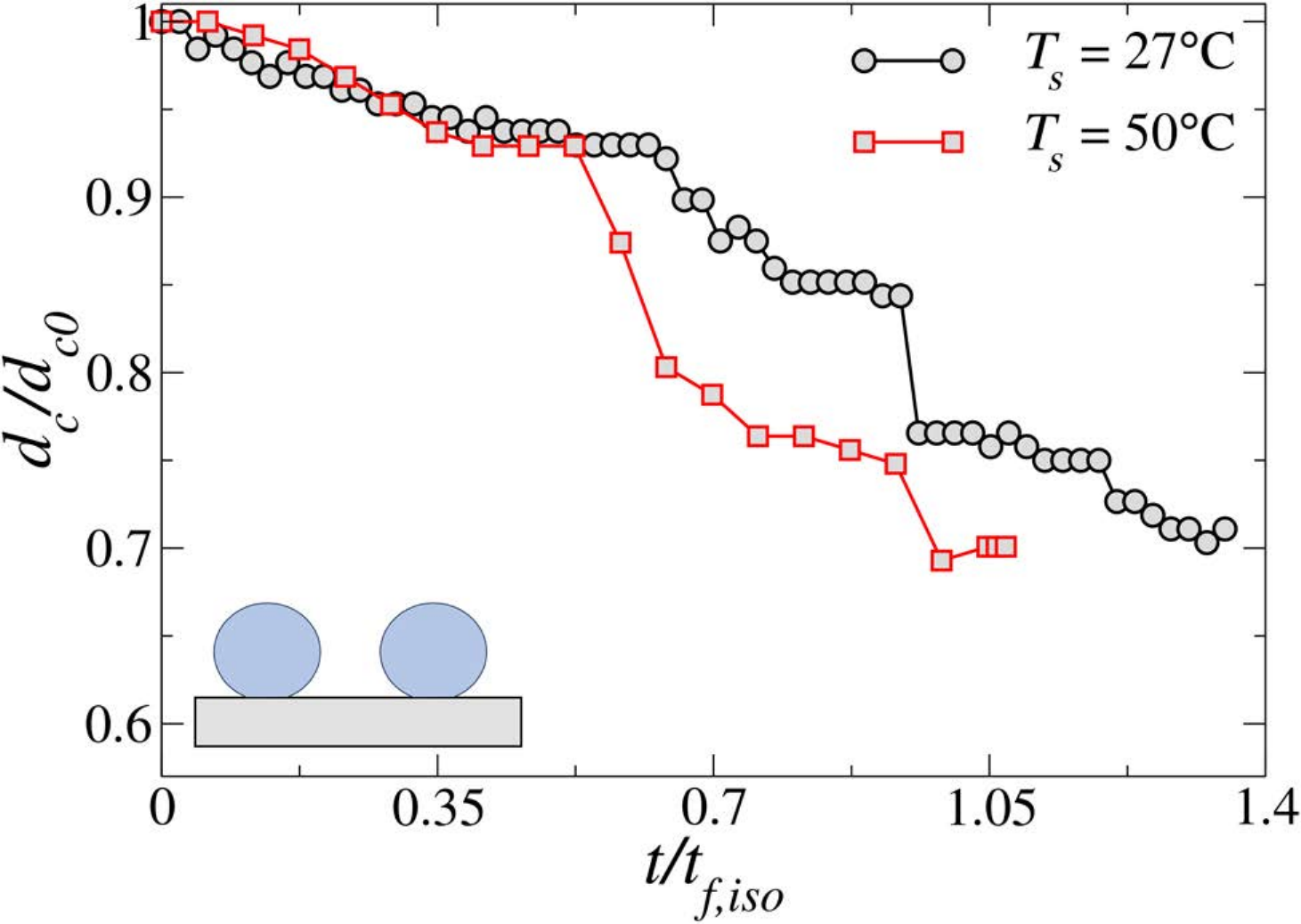} \hspace{2mm} \includegraphics[width=0.46\textwidth]{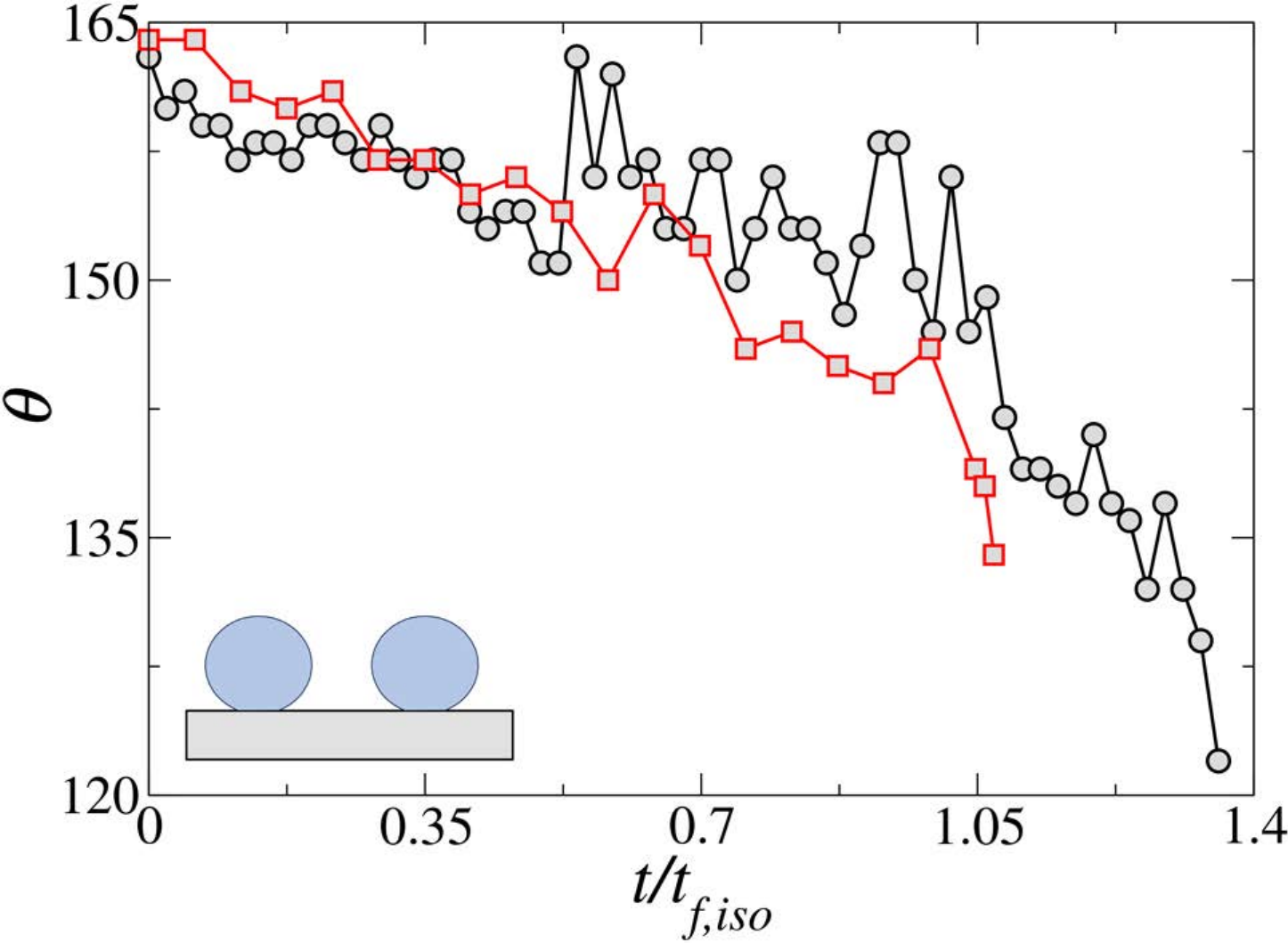} \\
Figure S4: {Temporal variation of the (a,c) normalized wetting diameter ($d_c/d_{c0}$) and (b,d) contact angle ($\theta$) at different substrate temperatures. Panels (a,b) correspond to the isolated droplet configuration, while panels (c,d) correspond to the two-drop configuration. The total evaporation time of the isolated droplet, denoted by $t_{f,\mathrm{iso}}$, is 1280 for $T_s = 27^\circ$C and 498 for $T_s = 50^\circ$C.}
\label{fig:figS4}
\end{figure}

\clearpage

\section{Calculation of the correction factor $K(E,\theta)$}

The correlations for the correction factor $K(E,\theta)$ are derived from the data provided by \citet{erbil2023droplet}, \citet{shen2022numerical}, and \citet{jenkins2023suppression}, and they are valid for contact angles up to $170^\circ$. The value of $K$ for $E$ between 0.1 and 1.0 is calculated by interpolation.
\begin{equation}\tag{S1}
\begin{split}
K(0.1,\theta) = & -2.55864 \times 10^{-11} \theta^5 + 9.29409 \times 10^{-9} \theta^4 - 1.3473 \times 10^{-6} \theta^3 \\
& + 7.44626 \times 10^{-5} \theta^2 - 0.00208963 \theta + 1.00388.
\end{split}
\end{equation}

\begin{equation}\tag{S2}
\begin{split}
K(1.0,\theta) = & -1.635853 \times 10^{-11} \theta^5 + 6.827309 \times 10^{-9} \theta^4 - 9.860522 \times 10^{-7} \theta^3 \\
& + 5.371728 \times 10^{-5} \theta^2 - 0.006027799 \theta + 1.000355.
\end{split}
\end{equation}

\end{document}